\documentclass[12pt]{article}
\usepackage{epsfig,amsmath,amssymb}

  {\end{list}}%

\makeatletter
\renewcommand{\section}{\setcounter{equation}{0}\@startsection
 {section}%
 {1}%
 {0pt}%
 {-1\baselineskip}%
 {0.4\baselineskip}%
 {\bfseries\large}}%
\renewcommand{\subsection}{\@startsection
 {subsection}%
 {2}%
 {0pt}%
 {-0.75\baselineskip}%
 {0.2\baselineskip}%
 {\bfseries}}%
\renewcommand{\subsubsection}{\@startsection
 {subsubsection}%
 {3}%
 {0pt}%
 {-0.5\baselineskip}%
 {0.1\baselineskip}%
 {\sc}}%
\makeatother

\def\a{\alpha}
\def\b{\beta}

\def\ga{\gamma}
\def\g5{\gamma_{5}}

\def\la{\lambda}

\def\ka{\kappa}

\def\r{\rho}
\def\s{\sigma}

\def\th{\theta}
\def\thb{\bar{\theta}}
\def\eps{\epsilon}

\def\om{\omega}

\def\ad{\dot{\alpha}}
\def\bd{\dot{\beta}}

\def\rd{\dot{\rho}}
\def\sd{\dot{\sigma}}

\def\thb{\bar{\th}}
\def\psib{\bar{\psi}}

\def\lab{\bar{\la}}
\def\sb{\bar{\sigma}}
\def\epsb{\bar{\eps}}

\def\prslash{{\partial\mkern-9mu/}}

\def\prslash{{\partial\mkern-9mu/}}    

\def\hdeps{\widehat\delta_{\epsilon}}

\def\idx{\int\!\! d^4\!x}

\newcommand{\bea}{\begin{eqnarray}}
\newcommand{\eea}{\end{eqnarray}}
\newcommand{\beann}{\begin{eqnarray*}}
\newcommand{\eeann}{\end{eqnarray*}}
\newcommand{\ba}{\begin{array}}
\newcommand{\ea}{\end{array}}

 \def\g {\gamma}

\begin{document}
 \begin{titlepage}
\rightline{FTI/UCM 133-2008}
\vglue 33pt

\begin{center}

{\Large \bf The Seiberg-Witten map and supersymmetry}\\
\vskip 1.0true cm
{\rm C. P. Mart\'{\i}n}\footnote{E-mail: carmelo@elbereth.fis.ucm.es}
 and C. Tamarit\footnote{E-mail: ctamarit@fis.ucm.es}
\vskip 0.1 true cm
{\it Departamento de F\'{\i}sica Te\'orica I,
Facultad de Ciencias F\'{\i}sicas\\
Universidad Complutense de Madrid,
 28040 Madrid, Spain}\\
\vskip 0.85 true cm

{\leftskip=50pt \rightskip=50pt \noindent
The lack of any local solution to the first-order-in-$h\omega^{mn}$ Seiberg-Witten (SW) map equations
for $U(1)$ vector superfields compels us to obtain the most general solution to those equations that
is a quadratic polynomial in the ordinary vector superfield,$v$, its chiral and  antichiral projections and the
susy covariant derivatives of them all. Furnished with this solution, which is local in the susy Landau gauge,  we construct  an ordinary dual of noncommutative $U(1)$ SYM in terms of ordinary fields which  carry a linear representation of the
${\cal N}=1$ susy algebra.
By using the standard SW map for the ${\cal N}=1$ $U(1)$ gauge supermultiplet we
define an ordinary $U(1)$ gauge theory which is dual to noncommutative $U(1)$ SYM in the WZ gauge. We show that the ordinary
dual so obtained is supersymmetric, for, as we prove as we go along,  the ordinary gauge and  fermion fields that we use to define it carry
a nonlinear representation of the ${\cal N}=1$ susy algebra. We finally show that the two ordinary duals of noncommutative $U(1)$ SYM introduced above are actually the same ${\cal N}=1$ susy gauge theory. We also show in this paper that the standard SW map is never the $\theta\bar{\theta}$ component of a local superfield in $v$ and check that, at least at a given approximation, a suitable field redefinition of that map makes the noncommutative and ordinary --in a $B_{mn}$ field-- susy $U(1)$ DBI actions equivalent.\par}
\end{center}

\vspace{9pt}
\vfill
\end{titlepage}

\section{Introduction}

 Noncommutative quantum field theories have been widely
investigated in the past years, chiefly after it was shown in
ref.~\cite{Seiberg:1999vs} that they arise as effective theories of open
strings ending on D-Branes with a constant Neveu-Schwarz  background
$B_{mn}$. In ref.~\cite{Seiberg:1999vs}, it was also  shown that  noncommutative
U(1) gauge theories can be mapped to a theory with ordinary gauge symmetry,
since both theories arise as  effective theories of the same underlying open string theory;
this equivalence can be seen~\cite{Seiberg:1999vs} as a mapping between a noncommutative Moyal
deformed DBI action and a commutative DBI action in the presence of a constant $B_{mn}$ background.
The Seiberg-Witten map thus associates to every noncommutative $U(N)$ gauge theory an equivalent --at least for energies well below the noncommutative energy scale-- ordinary $U(N)$ gauge theory, which we shall call in the sequel the ordinary dual under the Seiberg-Witten map of the former noncommutative gauge theory.

Most of the papers --see refs.~\cite{Liu:2000mja,Bichl:2001yf,Kraus:2001xt,Picariello:2001mu, Banerjee:2001un, Ghosh:2002wd, Grimstrup:2003rd, Banerjee:2004rs, Vilar:2006xk, Stern:2008wi, Schupp:2008fs} and~\cite{Marculescu:2008gw} for an  incomplete list-- where the properties of the ordinary duals under the Seiberg-Witten map of noncommutative $U(N)$ gauge theories are discussed deal with nonsupersymmetric theories or with the bosonic sector of supersymmetric theories. The construction
of supersymmetric duals under the Seiberg-Witten map of noncommutative supersymmetric $U(N)$ gauge theories is tackled only in an astonishingly
short number of papers --see for instance refs.~\cite{Ferrara:2000mm, Chekhov:2001ut, Paban:2002dr, Putz:2002ib, Dayi:2003ju, Mikulovic:2003sq, UlasSaka:2007ue}. Moreover the picture emerging from them is a bit blurred since there are important issues that have not been  clarified in them and
which we shall spell  out next.
First, there is the issue of the existence of a generalisation to objects made out of superfields of  the Seiberg-Witten map introduced in ref.~\cite{Seiberg:1999vs} --the map in ref.~\cite{Seiberg:1999vs} will be called henceforth the standard Seiberg-Witten map.
 In refs.~\cite{Ferrara:2000mm} and~\cite{Dayi:2003ju} it is claimed that there exists  such a generalisation and that it is a polynomial --and thus a local object-- in the ordinary vector superfield and its supersymmetry covariant derivatives. This statement is at odds with the result presented in ref.~\cite{Mikulovic:2003sq} where it is shown that
the first-order consistency condition for the Seiberg-Witten map for superfields admits no solution that is a polynomial of the appropriate ordinary superfields and their supersymmetry covariant derivatives. The latter result is in line with the fact that no local solution to the Seiberg-Witten map equations was found in ref.~\cite{Chekhov:2001ut} at first order in the noncommutativity parameter and with
the claim made in ref.~\cite{Putz:2002ib} that there is no superfield formalism in terms of ordinary vector superfields that would allow us to formulate the ordinary dual under the standard Seiberg-Witten map of noncommutative $U(1)$ superYang-Mills theory in the Wess-Zumino gauge. In
ref.~\cite{Chekhov:2001ut} a solution to the Seiberg-Witten map equations for $U(1)$ superfields was worked out at first order in the noncommutativity
parameter. In ref.~\cite{Chekhov:2001ut},  it is also claimed that the solution displayed in there is unique, which is quite surprising. The Seiberg-Witten map obtained by the authors of
ref.~\cite{Chekhov:2001ut} is local and trivial in the supersymmetric Landau gauge --but nonlocal and non-trivial otherwise-- and yields an ordinary  dual with linearly realised supersymmetry of the noncommutative $U(1)$ ${\cal N}=1$ superYang-Mills theory. Secondly, there is the issue of the supersymmetric character of the ordinary dual under the standard Seiberg-Witten map of $U(1)$ superYang-Mills theory in the Wess-Zumino gauge.
Such ordinary dual theory is constructed in refs.~\cite{Putz:2002ib, Dayi:2003ju} and~\cite{UlasSaka:2007ue}. In these  papers, the transformations of the fields of the dual ordinary theory that give rise to the supersymmetry transformations of the noncommutative theory are computed at first-order in the noncommutativity parameter. Those transformations of the dual ordinary fields  turn out to be nonlinear, though local, in these fields. It is thus apparent  that these  ordinary dual fields do not carry a linear
realisation of the ${\cal N}=1$ supersymmetry algebra in four dimensions. Whether these nonlinear transformations constitute a nonlinear realisation of the ${\cal N}=1$ supersymmetry algebra in four dimensions is not discussed in those papers, although the transformations in question are referred to as supersymmetry transformations. We believe that to rightly call these transformations supersymmetry transformations one should establish first that they are nonlinear realisations of the supersymmetry algebra. It should also be noticed that in general the Seiberg-Witten map does not preserve the gauge-fixing condition --e.g., it does not map in general an ordinary gauge field configuration in the temporal gauge into a noncommutative gauge field configuration in that very gauge, a situation that is reproduced for the Wess-Zumino gauge for the superfield Seiberg-Witten map of  ref.~\cite{Chekhov:2001ut}-- so it is not obvious that by choosing the Wess-Zumino gauge and then applying the standard Seiberg-Witten map one does not gives rise to a breaking of supersymmetry in the ordinary dual theory so constructed. Now that there seem to arise two ordinary duals --one obtained by using a nonlocal superfield Seiberg-Witten map and the other constructed by using the standard Seiberg-Witten map-- of   noncommutative $U(1)$ ${\cal N}=1$ superYang-Mills theory, it is fair to ask whether they  really are different theories as ordinary theories --they seem to have different supersymmetric features-- or the same ordinary theory expressed in terms of different sets of field variables. We have just stated the third issue that has not been clarified yet. Let us mention that in gaining a full understanding of all these matters one should  check --a check that has not been done in the literature yet-- that the standard Seiberg-Witten map, or some Seiberg-Witten map equivalent to it, establishes a connection between the ordinary DBI action in the presence of a constant background $B_{mn}$ field and the noncommutative DBI action for ${\cal N}=1$ supersymmetry in four dimensions.

The purpose of this paper is to clarify all the issues commented upon above. Before we display how we have organised the paper, let us point out
that a complete understanding of the duality relationship established by the Seiberg-Witten map for noncommutative supersymmetric gauge theories at the classical level is necessary, if the existence of such duality relationship is to be investigated for quantum theories --only for Chern-Simons theory such investigation has been undertaken~\cite{Kaminsky:2003qq}. Indeed, on the one hand, due to UV/IR mixing,  noncommutative non-supersymmetric Yang-Mills theories have severe noncommutative infrared divergences that are absent in their
supersymmetric versions~\cite{Ruiz:2000hu}, and, on the other hand, the ordinary dual theory of a given noncommutative gauge theory --e.g., noncommutative QED-- is not necessarily renormalisable~\cite{Wulkenhaar:2001sq}. We would also like to  stress that the  results we shall report on below are also relevant to the field of noncommutative gauge theories constructed within the enveloping-algebra formalism. This formalism  was
put forward in refs.~\cite{Madore:2000en, Jurco:2000ja} and~\cite{Jurco:2001rq}, and has led to important new results such as the formulation of the noncommutative Standard Model~\cite{Calmet:2001na} and other~\cite{Aschieri:2002mc} anomaly free theories~\cite{Brandt:2003fx}, which may be of
relevance in accounting for the experimental data to be recorded at the LHC ~\cite{Behr:2002wx,Melic:2005hb,Alboteanu:2006hh, Buric:2007qx}.

The layout of this paper is as follows. In Section 2 we show by explicit computation that the standard Seiberg-Witten map is never the $\theta\bar{\theta}$ component of a superfield made out of the ordinary vector superfield and its supersymmetry covariant derivatives, and
address the problem of finding physically sensible solutions to the Seiberg-Witten map equation for $U(1)$ superfields. Here, we
construct, at first order in the noncommutativity parameter, the most general solution --which is not unique-- to this equation  that is a quadratic polynomial in the ordinary vector superfield, its chiral and antichiral projections and their supersymmetry covariant derivatives. We show in Section 3 that the standard Seiberg-Witten map, applied to the noncommutative gauge supermultiplet of noncommutative $U(N)$ superYang-Mills theory in the Wess-Zumino gauge, always yields an ordinary $U(N)$ gauge supermultiplet which carries a nonlinear representation of the ${\cal N}=1$ supersymmetry
algebra in four dimensions. We discuss here how this result is in agreement with the fact that, upon adding certain field redefinitions  --that we compute in Appendix B-- the standard Seiberg-Witten map turns, in some approximation, the ${\cal N}=1$ supersymmetric  DBI action in the presence of a $B_{mn}$ field into the noncommutative ${\cal N}=1$ supersymmetric  DBI action. In Section 4 we show that the dual ordinary theories of
noncommutative $U(1)$ superYang-Mills theories constructed in Sections 1 and 2 are the same supersymmetric theory but formulated in terms
of different sets of variables. Our summary of the paper and the conclusions are the content of Section 5. We also include three appendices. Appendix A is merely notational. In
Appendix B we discuss the equivalence under the Seiberg-Witten map of the  DBI action in the presence of a $B_{mn}$ field and the
noncommutative DBI action, in the case of ${\cal N}=1$ supersymmetry in four dimensions. We have included in Appendix C the proof that the ordinary dual under the standard Seiberg-Witten map of noncommutative $U(1)$ Yang-Mills theory cannot be turned into a supersymmetric theory by including
 in the action new local terms of the appropriate dimension, if the fields in the resulting  action carry a linear representation of ${\cal N}=1$ supersymmetry in four dimensions.

\section{The Seiberg-Witten map equation for superfields and an ordinary dual of
noncommutative $U(1)$ ${\cal N}=1$ superYang-Mills}

       The aim of this section is to obtain a $U(1)$ ordinary theory
with linearly realised ${\cal N}=1$ supersymmetry which is dual,
at least classically, to   noncommutative $U(1)$ ${\cal N}=1$ superYang-Mills. To do so we shall set up the Seiberg-Witten-map equations for $U(1)$ superfields and then build solutions to them. We shall also show that
these solutions cannot be  constructed by following the strategy suggested in
ref.~\cite{Ferrara:2000mm}.

We define noncommutative gauge theories with linearly realised ${\cal N}=1$
supersymmetry in terms of superfields as in refs.~\cite{Chu:1999ij,Terashima:2000xq}. Our superspace conventions will be those found in
ref.~\cite{FigueroaO'Farrill:2001tr} and
the Moyal product, ``$\star$'', of $a$ and $b$ will be given by $a\star b=a \exp\Big(\frac{ih}{2}\overleftarrow{\partial_m}\omega^{mn}\overrightarrow{\partial_n}\Big)b$;
$h$ sets the noncommutative scale. All along this paper, we will denote
space-time indices with Latin letters and spinor indices with Greek
letters. $V$  shall denote a $U(1)$ noncommutative vector superfield. Under noncommutative
$U(1)$ transformations  --defined by the chiral superfield $\Lambda$--
$V$  transforms as follows:
\begin{equation}
 e_\star^{V'}=e_\star^{i\bar\Lambda}\star
e_\star^V\star e_\star^{-i\Lambda},
\label{NCgauge}
\end{equation}
$e_\star^A$ denotes the exponential of $A$ defined in terms of the
usual power series with products replaced by star products. $\bar\Lambda$
is the conjugate of $\Lambda$.

Let $s_{nc}$ denote the  operator generating the
noncommutative BRS transformations of the superfields $V$, then,
eq.~\eqref{NCgauge} leads to
\begin{equation*}
 s_{nc}\,V =-\frac{i}{2}\,L_V(\bar{\Lambda}+\Lambda)+\frac{i}{2}\,L_V
\text{coth}_\star\Big(\frac{L_V}{2}\Big)(\bar{\Lambda}-\Lambda),
L_V=[V,\quad]_\star,
s_{nc}\,\Lambda=i\Lambda\star\Lambda,
\end{equation*}
where $\Lambda$ now denotes an infinitesimal Grassmann chiral superfield.
Let $v$ and $\lambda$ denote, respectively, an ordinary $U(1)$ vector and an ordinary
 $U(1)$ ghost superfields.
In keeping with the ideas underlying the Seiberg-Witten map, to obtain an
ordinary theory dual of  noncommutative $U(1)$ ${\cal N}=1$ superYang-Mills, one should first express
the $U(1)$ noncommutative superfields $V$ and $\Lambda$
as functions of $v$ and $\lambda$, and
their susy covariant derivatives, in such a way that ordinary BRS orbits are mapped into
noncommutative BRS orbits. This is achieved by solving the Seiberg-Witten-map
equations for $U(1)$ superfields. These equations read
\begin{displaymath}
\begin{array}{l}
 s_{nc}\, \Lambda[\la,v]=s\Lambda[\la,v], \quad\quad\quad \Lambda,\lambda \quad\text{
 chiral},\\
 s_{nc}\,V[v]=sV[v],\quad\quad \quad V, v \quad\text{  real}.\\
\end{array}
\end{displaymath}
The symbol $s$ denotes the ordinary $U(1)$ BRS operator, which acts on the ordinary superfields as follows
\begin{equation}
sv=i(\bar{\la}-\la),\quad s\la=0.
\label{ORBRS}
\end{equation}

 Expanding the noncommutative fields in
powers of $h\omega^{mn}$,
\begin{equation}
 \Lambda=\la+h\Lambda^{(1)}+O(h^2),\quad
 V=v+hV^{(1)}+O(h^2),
 \label{firstordexp}
\end{equation}
one gets the following equations for the first order contributions:
\begin{align}
 \nonumber s\Lambda^{(1)}&=\frac{1}{32}\,\omega_{\a\b}\,\partial^{\dot{\ka}\a}\la
{\partial_{\dot{\ka}}}^\b
\la+\frac{1}{32}\,\omega_{\dot{\a}\dot{\b}}\,\partial^{\dot{\a}\ka}\la
{\partial^{\dot{\b}}}_\ka \la,\\
\label{SWeq}sV^{(1)}&=-\frac{1}{32}\,\omega^{\a\b}\,{\partial^{\dot{\ka}}}_\a
v\la \partial_{\dot{\ka}\b}
(\la+\bar{\la})-\frac{1}{32}\,\omega^{\dot{\a}\dot{\b}}\,{\partial_{\dot{\a}}}^\ka
v \partial_{\dot{\b}\ka}(\la+\bar{\la})+i(\bar{\Lambda}^{(1)}-\Lambda^{(1)}),
\end{align}
where we used --see Appendix A for notation-- the following relations between vector indices (Latin
letters) and spinor indices (Greek letters):
\begin{align*}
 \partial_{\dot{\a}\b}&=(\bar{\sigma}^m)_{\dot{\a}\b}\partial_m,\\
\omega^{mn}&=-\frac{1}{16}(\sigma^{mn})^{\a\b}\omega_{\a\b}+\frac{1}{16}(\bar{\sigma}^{mn})^{\dot{\a}\dot{\b}}\omega_{\dot{\a}\dot{\b}},\\
\omega^{\r\s}&=-2(\sigma^{mn})^{\r\s}\omega_{mn},\quad
\omega^{\dot{\r}\dot{\s}}=2(\bar{\sigma}^{mn})^{\dot{\r}\dot{\s}}\omega_{mn}.
\end{align*}
One should first look for solutions to eq.~\eqref{SWeq} that would allow us to make
contact with the Seiberg-Witten map --called the standard Seiberg-Witten map--
as introduced in ref.~\cite{Seiberg:1999vs}. In looking for these solutions the first obstacle one stumbles on
is the fact that, at first order in $h\omega^{m n}$, the standard Seiberg-Witten map is never the $\theta^{\a}\bar{\theta}^{\dot{\alpha}}$ component of a real superfield, with no free spinor indices, which is a  polynomial in $v$ and its susy covariant derivatives
$ D_\a,\bar D_{\dot \a},\partial_{\a\dot{\beta}}$. This fact,  that has not  been properly discussed in the literature as yet, contradicts the claim made in ref.~\cite{Ferrara:2000mm}
that the standard Seiberg-Witten map can be supersymmetrised at first order in
$h\omega^{m n}$, i.e., that at first order in
$h\omega^{m n}$ a dimensionless real polynomial in $v$ and its susy derivatives with no free spinor indices can be constructed so that its  $\theta^{\a}\bar{\theta}^{\dot{\alpha}}$ component is the standard Seiberg-Witten map.

The  BRS transformations with nonstandard normalisations inherited by the gauge fields $A_m$ --noncommutative-- and $a_m$ --ordinary--
from the superfield gauge transformations in eqs.~\eqref{NCgauge} and ~\eqref{ORBRS} read
 $s_{nc} A_m=-2\partial_m Z-i[A_m,Z]_{\star}$, $sa_m=-2\partial_m z$. For these
BRS transformations the first-order-in-$h\omega^{mn}$ standard Seiberg-Witten map of ref.~\cite{Seiberg:1999vs} runs thus
\begin{equation}
 A_l^{(1)\rm st}=\frac{1}{2}\om^{mn}\Big(a_m\partial_n
a_l-\frac{1}{2}a_m\partial_l a_n\Big).
\label{Ast}
\end{equation}
Let us now show that this $ A_l^{(1)\rm st}$ is not the
$\theta^{\beta}\bar{\sigma}_{\dot{\alpha}\beta}\bar{\theta}^{\dot{\alpha}}$ component of a dimensionless real polynomial with no free spinor  indices
made out of $v$ and its susy derivatives. Since $ A_l^{(1)\rm st}$ is quadratic in $a_m$, it suffices to consider the most general, $\tilde V$, dimensionless real polynomial in $v$ and its susy derivatives with no free spinor indices which
is linear in $\omega^{mn}$ and quadratic in $v$. $\tilde V$ is given by
\begin{align*}
 \tilde V=\sum_{i=1}^5 (x_i\,{\rm Re}\,t_i\,+\,y_i\,{\rm Im}\,t_i),\,x_i,y_i\in\mathbb{R},
\end{align*}
where ${\rm Re}\,t_i$ and ${\rm Im}\,t_i$ denote, respectively, the real and imaginary parts of $t_i$, and
$x_i$ and $y_i$ are arbitrary real coefficients. $\{t_i\}_{\{i=1\dots 5\}}$ denotes the following set of monomials
\begin{equation}
\begin{array}{l}
{ t_1=\omega^{\a\b}\partial_{\a\dot\b}\bar D^{\dot\b}D_\b  v
v,\; t_2=\omega^{\a\b}\partial_{\a\dot\b}\bar D^{\dot\b}vD_\b v, \;
t_3=i\omega^{\a\b}\bar D^2 D_\a v D_\b v,\;
t_4=\omega^{\a\b}{\partial_\a}^{\dot\b} D_\b v \bar D_{\dot\b}v,}\\
{t_5=\omega^{\a\b}D_\a\bar D^{\dot\a}v\partial_{\b\dot\a}v.
\label{rs}}\\
\end{array}
\end{equation}
For the reader's sake we also display the complex conjugates, $\bar{t}_i,\,i=1\dots 5$, of the previous monomials:
\begin{displaymath}
\begin{array}{l}
{ \bar{t}_1\!=\!-\omega^{\dot{a}\dot{\b}}\partial_{\b\dot{\a}} D^{\b}\bar{ D}_{\dot{\b}}  v
v,\;\bar{t}_2\!=\!-\omega^{\dot{\a}\dot{\b}}\partial_{\b\dot{\a}} D^{\b}v\bar D_{\dot{\b}} v,\;
\bar{t}_3\!=\!-i\omega^{\dot{\a}\dot{\b}} D^2\bar D_{\dot{\a}} v\bar D_{\dot{\b}} v,\;
\bar{t}_4\!=\!-\omega^{\dot{\a}\dot{\b}}{\partial^{\b}}_{\dot\a}\bar D_{\dot\b} v  D_{\b}v,}\\
{\bar{t}_5\!=\!-\omega^{\dot\a\dot\b}\bar D_{\dot \a} D^{\a}v\partial_{\a\dot\b}v.}
\end{array}
\end{displaymath}
Let us now show that for  no choice of $x_i$ and $y_i$ the following equation will hold
\begin{align}
 \frac{1}{4}(\sigma_m)^{\b\dot\a}[\bar D_{\dot\a},D_\b]\tilde
V|_{\th=\bar\th=0,aa}=A^{(1)\rm st}_m.
\label{Astsup}
\end{align}
By $aa$, we mean that only the contributions quadratic in $a_m$ are kept. Now,
it can be seen that the $a_m-$dependent part of the terms ${\rm Im}\,t_i$ always involve
contractions with the Levi-Civita symbol $\eps^{mnrs}$, which never
occur in eq.~\eqref{Ast} --recall that $\omega^{mn}$ is real and that our noncommutative space-time has got Minkowski signature. Hence,  the $y_i$ will be of no avail to make  eq.~\eqref{Ast} hold and thus we shall only worry about the contributions coming from ${\rm Re}\,t_i$. Introducing the notation $\frac{1}{4}(\sigma_m)^{\b\dot\a}[\bar
D_{\dot\a},D_\b]{\rm Re}\,t_i\equiv \tilde A^{
[i]}_m$ and after some computations one  finds that
\begin{align*}
\tilde A^{[1]}_l&=-4\,\omega^{mn}f_{mn}a_l,\\
\tilde A^{[2]}_l&=-2\,\omega^{mn}(f_{mn}a_l+2\eta_{ml}a_n(\partial
a)+2f_{ml}a_n-2\eta_{ln}f_{mk}a^k),\\
\tilde A^{[3]}_l&=-8\,\omega^{mn}(f_{mn}a_l-2f_{lm}a_n-2\eta_{lm}f_{kn}a^k),\\
\tilde A^{[4]}_l&=-2\,\omega^{mn}(f_{mn}a_l-2f_{ln}a_m+4\partial_m a_l
a_n -2\eta_{ln}f_{mk}a^k+4\eta_{lm}\partial^ka_n
a_k+2\eta_{ln}(\partial a)a_m),\\
\tilde A^{[5]}_l&=16\,\omega^{mn}a_m\partial_n a_l.
\end{align*}
Finally, eq.~\eqref{Astsup} boils down  to
\begin{equation*}
 \sum x_i \tilde  A^{[i]}_l=\frac{1}{2}\om^{mn}\Big(a_m\partial_n
a_l-\frac{1}{2}a_m\partial_l a_n\Big),
\end{equation*}
which has no solution since, in spite of the fact that the terms that occur on its r.h.s. can be obtained
by choosing  several values of the $x_i$, there always appear
undesired extra terms involving contractions of the type
$\omega^{mn}\eta_{lm}$. Notice that the ambiguity~\cite{Asakawa:1999cu} of the Seiberg-Witten map
cannot be taken advantage of to fix this situation, for this ambiguity, in the $U(1)$ case, is linear in $a_m$.

In searching for solutions to  eq.~\eqref{SWeq}, the second difficulty one meets is that, as shown in ref.~\cite{Mikulovic:2003sq}, $\Lambda^{(1)}$
cannot be a polynomial in $v$, $\lambda$ and its susy derivatives, since $\Lambda^{(1)}$ is chiral. Thus one is led to look for nonlocal solutions to  eq.~\eqref{SWeq}, i.e., solutions that
are not polynomials in the ordinary superfields and their susy derivatives.
To avoid the inconsistencies that usually arise in theories with gauge
independent nonlocal terms, one may look for solutions to eq.~\eqref{SWeq}
whose nonlocal contributions vanish in a given gauge. Since both the
chiral and antichiral projections of $v$, namely, $ v_+\equiv P_+v$ and
$ v_-\equiv P_-v$,  with $P_+=\frac{1}{16\pi^2}\frac{\bar D^2D^2}{\square}$ and $P_-=\frac{1}{16\pi^2}\frac{D^2\bar D^2}{\square}$, vanish in the susy
Landau gauge $D^2\bar D^2v=\bar D^2 D^2v=0$, and since   projecting $v$ into its
chiral part may help find a chiral $\Lambda^{(1)}$, it is natural --and the next simplest ansatz to that of local solutions-- to look for
solutions to  eq.~\eqref{SWeq} that are polynomials in $\la,v,v_+,v_-$
and their susy covariant derivatives. We shall further assume that $\Lambda^{(1)}$
is linear in $v,v_{\pm}$, and that $V^{(1)}$ is at most
quadratic in $v,v_{\pm}$; the rationale for these assumptions is that the corresponding first-order-in-$h\omega^{mn}$
contributions to the standard Seiberg-Witten maps are, respectively,
linear and quadratic in $a_m$. Let us introduce some more  notation:
$\bar v\equiv v-v_+-v_-$ --of course, $s\bar v=0$. A lengthy computation
yields the following family of solutions to eq.~\eqref{SWeq}:
\begin{align}
       \Lambda^{(1)}&=\frac{i}{32}\,\omega^{\a\b}\,{\partial^{\dot{\a}}}_\a
v_+\partial_{\dot{\a}\b}\la+\frac{i}{32}\,\omega^{\dot\a\dot\b}\,{\partial_{\dot{\a}}}^\a
v_+\partial_{\a\dot{\b}}\la+ x
\omega^{\a\b}\bar{D}^2(D_\a\bar{v}D_\b\la),\label{SWLambda}\\
V^{(1)}&= x \omega^{\a\b}\bar{D}^2(D_\a\bar{v}D_\b v)+
\bar{x}\omega^{\dot\a\dot\b}{D}^2(\bar{D}_{\dot{\a}}\bar{v}\bar{D}_{\dot{\b}}
v)\label{SWV}\\ 
\nonumber&+\frac{i}{32}\,\omega^{\a\b}\,[{\partial^{\dot{\a}}}_\a(v-v_-)\partial_{\dot{\a}\b}(v-v_+)]-\frac{i}{32}\,\omega^{\dot\a\dot\b}\,[{\partial_{\dot{\a}}}^\a(v-v_+)\partial_{\dot{\b}\a}(v-v_-)]+{\cal
X},\quad s {\cal X}=0.
\end{align}
$x$ is an arbitrary constant parametrising the ambiguity in the map
for $\Lambda$; it must be imaginary if one wants to avoid --as happens in standard
Seiberg-Witten map case-- parity
violating terms --contributions involving contractions with the
$\epsilon^{mnrs}$ symbol-- in the map for the component field $a_m$ that otherwise will make the noncommutative and ordinary gauge fields  behave not in the same way under parity.
${\cal X}$ represents the ambiguity in the map for the real superfield $V$;
it is given by the most general linear combination of terms constructed from
$\bar v$ and susy covariant derivatives $D_\a,\bar D_{\dot\a },\partial_{\dot\a\b}$,
i.e., a linear combination of the real and imaginary parts of the
terms appearing in eq.~\eqref{rs}, with $v$ substituted by $\bar{v}$.
${\cal X}$ can be interpreted as a field redefinition of $v$.
Our solutions for $\Lambda^{(1)}$ and $V^{(1)}$ include the particular
solution found in ref.~\cite{Chekhov:2001ut}.

In the case of the map for $V,$ the $x$-dependent terms can be
gauged away by  performing a gauge transformation of $v$, since they can be written as the
difference of a chiral and an antichiral term. It is plain that in the supersymmetric Landau gauge
the Seiberg-Witten map above is local and $V^{(1)}$ is given by the most general
local expression quadratic in $v$ that one can write; this is a very
welcomed feature of the map in regards with renormalisability issues~\cite{Buric:2006wm, Martin:2007wv, Buric:2007ix}.

In refs.~\cite{Putz:2002ib, Dayi:2003ju} the standard Seiberg-Witten map was used to construct
an ordinary --i.e., on ordinary Minkowski space-time-- field theory that is dual to $U(1)$ noncommutative SYM theory formulated in the Wess-Zumino gauge. This ordinary dual theory is formulated in terms of the  ``susy" gauge multiplet $(a_m,\lambda_{\a},d)$, which undergoes ordinary $U(1)$ transformations but whose ``susy" transformations
are a sum of the ordinary susy transformations plus nonlinear $\omega^{mn}$-dependent terms --this is why for the time being we write
``susy" and not susy; we shall show that these comas can be removed in Section 3. Since it is one of the purposes of this paper to relate the ordinary dual
theory obtained from noncommutative $U(1)$ ${\cal N}=1$ superYang-Mills by using the Seiberg-Witten map for superfields --see eqs..~\eqref{firstordexp},~\eqref{SWLambda} and~\eqref{SWV}-- with the dual ordinary theory obtained from the latter noncommutative theory as in refs.~\cite{Putz:2002ib, Dayi:2003ju}, we shall need to gauge transform to the Wess-Zumino gauge the noncommutative scalar superfield $V[v]$ defined in eqs.~\eqref{firstordexp} and~\eqref{SWV}. Let us stress first that if
$v^{\rm WZ}$ denotes a general ordinary real scalar superfield in the Wess-Zumino gauge, then its noncommutative image,
$V[v^{\rm WZ}]$, given by the Seiberg-Witten map in eqs.~\eqref{firstordexp} and~\eqref{SWV}, is not a noncommutative real scalar superfield in the Wess-Zumino gauge. But, of course, one can further gauge transform this $V[v^{WZ}]$
to a new noncommutative scalar superfield $V^{\rm WZ}[a_m,\lambda_{\a},\bar \lambda_{\dot\a}, d]$ which is in the Wess-Zumino gauge --$a_m,\lambda_{\a},\bar \lambda_{\dot\a}$ and $d$ are the components of $v^{\rm WZ}$. Indeed,
\begin{equation}
e^{V^{WZ}[a_m,\lambda_{\a},\bar \lambda_{\dot\a}, d]}=e_\star^{i\bar{\Lambda}^{\rm WZ}}\star
e_\star^{V[v^{\rm WZ}]}\star e_\star^{i\Lambda^{\rm WZ}},
\label{gaugingtoWZgauge}
\end{equation}
for a  $\Lambda^{\rm WZ}$ which is linear in $h\omega^{mn}$, leads to
\begin{align}
\nonumber V^{\rm WZ}[a_m,\lambda_{\a},\bar \lambda_{\dot\a}, d]&=v^{\rm WZ}+hV^{(1)}[v^{\rm
WZ}]+ih(\bar{\Lambda}^{\rm WZ}-\Lambda^{\rm WZ})+O(h^2),\\
\Lambda^{\rm WZ}&=-\frac{i}{2}C^{(1)}(y)-i\th^\a\Psi^{(1)}_\a(y)-\frac{i}{2}\th^2F^{(1)}(y),
y^m=x^m-i\th^\a\sb^m_{\bd\a}\thb^{\bd}.
\label{VWZ}
\end{align}
$C^{(1)}(x)$, $\Psi^{(1)}_\a(x)$ and $F^{(1)}(x)$ are the lowest components of $V^{(1)}[v^{\rm
WZ}]$, the latter defined by eq.~\eqref{SWV}:
\begin{equation}
\begin{array}{l}
{V^{(1)}[v^{\rm
WZ}]= C^{(1)}\!+\!\theta^{\a}\Psi^{(1)}_\a\!+\!\bar\theta^{\ad}\bar\Psi^{(1)}_{\ad}\!+\!\frac{1}{2}\theta^2 F^{(1)}
\!+\!\frac{1}{2}\bar\theta^2 \bar F^{(1)}\!+\!
\theta^{\a}\bar \theta^{\dot\beta}A^{(1)}_{{\dot\beta}\alpha}\!+\!
\frac{1}{2}\theta^2\bar\theta^{\dot\alpha}\bar\Lambda^{'\,(1)}_{\dot\alpha}\!+\!
\frac{1}{2}\bar \theta^2\theta^{\alpha}\Lambda^{'\,(1)}_{\alpha}}\\
{\phantom{V^{(1)}[v^{\rm WZ}]=}\!+\!\frac{1}{4}\bar\theta^2\theta^2 D^{'\,(1)},
}\\[4pt]
{\Lambda^{'\,(1)}_{\alpha}=\Lambda^{(1)}_{\alpha}\!-\!i
\sigma^{m}_{{\dot\beta}\alpha}\partial_m
\bar\Psi^{(1)\,\dot\beta},\quad  D^{'\,(1)}=D^{(1)}\!+\!\Box C^{(1)},}\\
{v^{WZ}=\theta^{\a}\bar \theta^{\dot\beta}a_{\dot{\b}\a}+\frac{1}{2}\theta^2\bar\theta^{\dot\alpha}\bar\lambda_{\dot\alpha}+
\frac{1}{2}\bar \theta^2\theta^{\alpha}\lambda_{\alpha}+\frac{1}{4}\bar\theta^2\theta^2 d.}
\label{compexpan}
\end{array}
\end{equation}
For $x=0$ and ${\cal X}=0$, the components of $V^{(1)}[v^{\rm
WZ}]$ read
\begin{align}
 \nonumber C^{(1)}=&-\frac{\om^{\a\b}}{256}\frac{\partial_\a^{\ad}}{\Box}d\frac{\partial_{\ad\b}}{\Box}\partial
a+c.c.,\\
\nonumber\Psi^{(1)}_\s=&\frac{\om^{\a\b}}{256}\frac{\partial_\a^{\ad}}{\Box}(d-2i\partial
a)\frac{\partial_{\ad\b}\partial_{\rd\s}}{\Box}\lab^{\rd}+\frac{\om^{\ad\bd}}{256}\frac{\partial_{\ad}^{\a}}{\Box}(d-2i\partial
a)\frac{\partial_{\bd\a}\partial_{\rd\s}}{\Box}\lab^{\rd},\\
\nonumber F^{(1)}=&0,\\
A^{(1)}_{\bd\ga}=&\frac{\om^{\a\b}}{256}\left[8\partial_\a^{\ad}
a_{\bd\g}\frac{\partial_{\ad\b}}{\Box}\partial
a+4\frac{\partial_\a^{\ad}}{\Box}\partial
a\frac{\partial_{\ad\b}\partial_{\bd\g}}{\Box}\partial
a+\frac{\partial_\a^{\ad}}{\Box}d\frac{\partial_{\ad\b}\partial_{\bd\g}}{\Box}d-2i\frac{\partial_\a^{\ad}\partial_{\bd\s}}{\Box}\la^\s
\frac{\partial_{\ad\b}\partial_{\g\sd}}{\Box}\lab^{\sd}\right]\label{V1components}\\
\nonumber&+(c.c)|_{(\b\leftrightarrow\gamma)},\\
\nonumber\Lambda^{(1)}_\r=&\frac{1}{128}\left[-4\om^{\a\b}\frac{\partial_\a^{\ad}}{\Box}\partial
a\partial_{\ad\b}\la_\r-4\om^{\ad\bd}\frac{\partial_{\ad}^{\a}}{\Box}\partial
a\partial_{\bd\a}\la_\r+2\omega^{\a\b}\frac{\partial^{\ad}_\a\partial^{\rd}_\s}{\Box}\la^\s\partial_{\ad\b}\Big(a_{\rd\r}-\frac{\partial_{\rd\r}}{\Box}\partial
a\Big)\right.+\\
\nonumber&+\left.2\omega^{\ad\bd}\frac{\partial^{\a}_{\ad}\partial^{\rd}_\s}{\Box}\la^\s\partial_{\bd\a}\Big(a_{\rd\r}-\frac{\partial_{\rd\r}}{\Box}\partial
a\Big)+i\omega^{\a\b}\frac{\partial_\a^{\ad}\partial_\s^{\rd}}{\Box}\la^\s\frac{\partial_{\ad\b}\partial_{\rd\r}}{\Box}d+i\omega^{\ad\bd}\frac{\partial_{\ad}^{\a}\partial_\r^{\rd}}{\Box}d\frac{\partial_{\bd\a}\partial_{\rd\s}}{\Box}\la^\s\right],\\
\nonumber D^{(1)}\!=&\frac{\om^{\a\b}}{128}\left[4\partial_{\a}^{\ad}d\frac{\partial_{\ad\b}}{\Box}\partial
a\!+\!\frac{\partial_\a^{\ad}\partial_\s^{\rd}}{\Box}\la^\s\partial_{\ad\b}\lab_{\rd}\!+\!
\partial_{\a}^{\ad}\la^\s\frac{\partial_{\ad\b}\partial_{\sd\s}}{\Box}\lab^{\sd}
\!+\!2\partial_\a^{\ad}\Big(a^{\s\rd}\!-\!\frac{\partial^{\s\rd}}{\Box}\partial
a\Big)\frac{\partial_{\ad\b}\partial_{\rd\s}}{\Box}d\right]\!\!+\!c.c.
\end{align}
       In the previous equations, $(c.c.)$ denotes complex conjugate and
$(c.c.)|_{\b\leftrightarrow\ga}$ denotes
complex conjugate with indices $\b$ and $\g$ exchanged (hermitian
conjugation); for example
$\s_{\bd\ga}+(c.c.)|_{\b\leftrightarrow\g}=2\s_{\bd\ga}.$

Taking into account eqs.~\eqref{VWZ},~\eqref{compexpan} and~\eqref{V1components}, one concludes that
\begin{equation}
\begin{array}{l}
{V^{\rm WZ}[a_m,\lambda_{\a},\bar \lambda_{\dot\a},d]=
\theta^{\a}\bar \theta^{\dot\beta}A_{\dot\beta\alpha}+
\frac{1}{2}\theta^2\bar\theta^{\dot\alpha}\bar\Lambda_{\dot\alpha}+
\frac{1}{2}\bar \theta^2\theta^{\alpha}\Lambda_{\alpha}
+\frac{1}{4}\bar\theta^2\theta^2 D,}\\[4pt]
{A_{{\dot\beta}\alpha}=a_{{\dot\beta}\alpha}+hA^{(1)}_{\dot \beta\alpha}+O(h^2),\quad
\Lambda_{\alpha}=\lambda_{\alpha}+h \Lambda^{(1)}_{\alpha}+O(h^2),\quad
D=d+hD^{(1)}+O(h^2),}
\end{array}
\label{nonlocalWZmap}
\end{equation}
where $A^{(1)}_{\dot \beta\alpha}$, $\Lambda^{(1)}_{\alpha}$ and $D^{(1)}$ are
the same as for $V[v^{\rm WZ}]$ and thus given in
eq.~\eqref{V1components}. Let us stress that $V[v^{\rm WZ}]$ and $V^{\rm WZ}[a_m,\lambda_{\a},\bar \lambda_{\dot\a},d]$ define the same theory since they are related by a noncommutative gauge transformation.

We shall close this section by recalling that the ambiguity ${\cal X}$ in the Seiberg-Witten map in eq.~\eqref{SWV} has no physical consequences since it is a local field redefinition of the ordinary vector superfield, hence we shall set it to zero from now on.

\section{Ordinary duals of noncommutative $U(N)$ ${\cal N}=1$ SuperYang-Mills
theory under the standard Seiberg-Witten map}

In refs.~\cite{Putz:2002ib} and~\cite{Dayi:2003ju}, the standard Seiberg-Witten
map was used to map noncommutative $U(1)$ SYM theory in the Wess-Zumino gauge
to an ordinary gauge theory with $U(1)$ symmetry. This construction can be
generalised to noncommutative $U(N)$ gauge groups as we shall do next. The construction we are about to develop
may be of relevance in studying some of the physical implications of the
models proposed in refs.~\cite{Chu:2001fe, Chaichian:2001py, Jaeckel:2005wt,  Abel:2005rh, Arai:2006ya} and~\cite{Arai:2007dm}.

Our supersymmetric noncommutative field theory will have the following field
content: a noncommutative gauge ${\cal N}=1$ supermultiplet, $(A_m,\Lambda_{\alpha},D)$. The fields $A_m,\Lambda_{\alpha},D$ are
valued in the Lie algebra of $U(N)$ in the fundamental representation. If $Z(x)=Z^a(x)\,{\rm T}^a$ denotes an infinitesimal function valued in the Lie algebra of $U(N)$ in the fundamental representation, with $Z^a(x)$ being ghost fields, our theory will be
invariant under the following noncommutative BRS transformations:
\begin{displaymath}
s^{nc}_{Z}A_m=-\hat D_m Z=-(\partial_m Z+i[A_m,Z]_{\star}), \quad
s^{nc}_{Z}\Lambda_{\alpha}=-i[\Lambda_{\alpha},Z]_{\star},\quad
s^{nc}_{Z}D=-i[D,Z]_{\star}.
\end{displaymath}
In addition to the BRS symmetry just defined, our $U(N)$ noncommutative  gauge theory
will be invariant under the following supersymmetry transformations:
\begin{equation}
\hdeps A_m=\frac{1}{4}\eps \s_m
\bar\Lambda-\frac{1}{4}\epsb\sb_m\Lambda,\quad
\hdeps \Lambda_\a=-\eps_\a
D+2i\eps_\ga{(\s^{mn})^\g}_\a F_{mn},\quad
\hdeps D=i\epsb\sb^m\hat D_m\Lambda+i\eps\s^m \hat D_m
\bar{\Lambda},
\label{ncsusytransf}
\end{equation}
where $F_{mn}=\partial_{m}A_n-\partial_{n}A_m+i[A_m,A_n]_{\star}$ and
$\hat D_m=\partial_{m}+i[A_m,\phantom{A_n}]_\star$.
These  supersymmetry transformations are linear modulo noncommutative gauge
transformations, hence the noncommutative multiplets of our theory carry a
linear representation of the supersymmetry algebra: of course, there is a
formulation of our theory in terms of superfields, each multiplet above
constituting the components of the appropriate superfield in the Wess-Zumino gauge.

Let $\tilde a_m$, $\tilde\lambda_{\alpha}$ and $\tilde d$ stand, respectively, for the ordinary
counterparts, under the standard Seiberg-Witten map, of the noncommutative fields
$A_m$, $\Lambda_{\alpha}$ and  $D$ introduced above.
Then, up to first order in $h\omega^{mn}$, the standard Seiberg-Witten map for our theory is given by the following equations
\begin{equation}
\begin{array}{l}
{A_m[\tilde a_n]=\tilde a_m+\frac{h}{4}\omega^{nl}\{\tilde a_n,\partial_l \tilde a_m+\tilde f_{lm}\}+O(h^2),}\\
{\Lambda_{\alpha}[\tilde a_m,\tilde\lambda_\alpha]=\tilde\lambda_{\alpha}+
\frac{h}{4}\omega^{mn}\{\tilde a_m,2 D_n\tilde\lambda_{\alpha}-i[\tilde a_n,\tilde\lambda_{\alpha}]\}+O(h^2),}\\
{D[\tilde a_m,\tilde d]=\tilde d+\frac{h}{4}\omega^{mn}\{\tilde a_m,2 D_n \tilde d-i[\tilde a_n,\tilde d]\}+O(h^2),
\label{standardWSmap}}\\
\end{array}
\end{equation}
where $\tilde f_{nl}=\partial_n \tilde a_l-\partial_l \tilde a_n+i[\tilde a_n,\tilde a_l]$, $D_m=\partial_m +i[\tilde a_m,\phantom{\tilde a_m}]$. By construction the Seiberg-Witten map
defined in eq.~\eqref{standardWSmap} maps infinitesimal gauge orbits of the
ordinary theory into infinitesimal gauge orbits of the noncommutative theory.
Indeed, if the noncommutative field ${\cal U}[\tilde a_m,{\it u}]$ is the image under the Seiberg-Witten map of ${\it u}$, then
\begin{equation}
{\cal U}[\tilde a_m,{\it u}]+\kappa\, s_{nc}\,{\cal U}[\tilde a_m,{\it u}]=
{\cal U}[\tilde a_m+\kappa\, \tilde s \tilde a_m, {\it u}+\kappa\, \tilde s{\it u}],
\label{aSWeq}
\end{equation}
$\kappa$ being the infinitesimal  BRS Grassmann parameter and $\tilde s$ being the
ordinary BRS operator which acts on our fields with tilde as follows:
\begin{displaymath}
\begin{array}{l}
{\tilde s_{z}\tilde a_m=-D_m z=-(\partial_m z+i[\tilde a_m,z]), \quad
\tilde s_{z}\Lambda_{\alpha}=-i[\Lambda_{\alpha},z],\quad
\tilde s_{z}\tilde d=-i[\tilde d,z].}\\
\end{array}
\end{displaymath}
Of course, in eq.~\eqref{aSWeq}, $Z$ in $s_{nc}$ and $z$ in
$\tilde s$ are not independent, but related by
\begin{equation}
Z=z+\frac{h}{4}\omega^{mn}\,\{\tilde a_{m},\partial_{n}z\}.
\label{SWmapforZ}
\end{equation}

We have seen that the Seiberg-Witten map in eq.~\eqref{standardWSmap} maps
a theory on ordinary space-time having an ordinary $U(N)$ gauge symmetry to
a noncommutative $U(N)$ gauge theory having, therefore, a noncommutative gauge
symmetry. But, this noncommutative gauge theory is further a supersymmetric
theory and its fields carry a linear --the supersymmetric transformations in
eq.~\eqref{ncsusytransf} are linear modulo noncommutative gauge transformations--
representation of the supersymmetry algebra, i.e., the commutator of two supersymmetry transformations acting
on a noncommutative field, ${\cal U}$, closes on space-time translations modulo a noncommutative gauge
transformation:
\begin{align}
 [\widehat\delta_\xi,\widehat\delta_\eta]\;{\cal U}(x)=-2i(\eta\sigma^m\bar\xi-\xi\sigma^m\bar\eta)\,\partial_m\;{\cal U}(x)+\delta^{(ncgauge)}_{\Omega}\;{\cal U}(x)\equiv
P\;{\cal U}(x)+\delta^{(ncgauge)}_{\Omega}\;{\cal U}(x).
\label{closureNC}
\end{align}
${\cal U}(x)$ denotes any of the noncommutative fields of our noncommutative theory. $\delta^{(ncgauge)}_{\Omega}\;{\cal U}(x)$ is a noncommutative gauge transformations with $\Omega(x)=
-2i(\eta\sigma^m\bar\xi-\xi\sigma^m\bar\eta)\,A_m (x)$. The next issue to be addressed is whether there
exist transformations of the ordinary fields that occur in the Seiberg-Witten map in
eq.~\eqref{standardWSmap} that give rise to the supersymmetry transformations of the corresponding noncommutative
fields that we have just discussed. The answer to this problem is that there exist such transformations
since we are dealing with $U(N)$ in the fundamental and antifundamental representations. Indeed,
we shall look for infinitesimal variations, $\tilde\delta_{\eps}{\it u}$, of the ordinary fields in eq.~\eqref{standardWSmap},
collectively denoted by ${\it u}$, such that
\begin{equation}
{\cal U}[\tilde a_m,{\it u}]+\hdeps\,{\cal U}[\tilde a_m,{\it u}]=
{\cal U}[\tilde a_m+\tilde\delta_{\eps}\,\tilde a_m, {\it u}+\tilde\delta_{\eps}{\it u}],
\label{ordsusyeq}
\end{equation}
where $\hdeps\,{\cal U}[\tilde a_m,{\it u}]$ is defined in eq.~\eqref{ncsusytransf}. Since we understand the Seiberg-Witten map
as a formal power series expansion in $h\omega^{mn}$, it turns out that $\tilde\delta_{\eps}{\it u}$ can be
obtained from eq.~\eqref{ordsusyeq} as a formal power series expansion in $h\omega^{mn}$, provided that the
representation of the gauge group that one considers satisfies: ${\rm L}_1\cdot {\rm L}_2$ belongs to its Lie algebra in the
corresponding representation, if ${\rm L}_1$ and ${\rm L}_2$ do. As  pointed out in ref.~\cite{Terashima:2000xq},
this condition restricts the type of gauge group to $U(N)$ groups, or products of them, and the type of irreducible
representation to the fundamental, antifundamental, adjoint and bi-fundamental. Up to first order in
$h\omega^{mn}$, we have
\begin{equation}
\begin{array}{l}
{\tilde\delta_{\eps} \tilde a_m= \frac{1}{4}\eps\sigma_m\bar{\tilde{\lambda}}-\frac{1}{4}\bar\eps\bar\sigma_m\tilde\lambda+\frac{h}{16}\omega^{nl}\;\Big[
                           \{\tilde a_n,2D_l(\eps\sigma_m\bar{\tilde{\lambda}}\!-\!\bar\eps\bar\sigma_{m}\tilde\lambda)\!
                           -\!i[\tilde a_l,\eps\sigma_m\bar{\tilde{\lambda}}\!-\!\bar\eps\bar\sigma_m\tilde\lambda]\}}\\
{\quad -\{\eps\sigma_n\bar{\tilde{\lambda}}\!-\!\bar\eps\bar\sigma_n\tilde\lambda,\partial_l \tilde a_m+\tilde f_{lm}\}\!
 -\!\{\tilde a_n,\partial_l (\eps\sigma_m\bar{\tilde{\lambda}}\!-\!\bar\eps\bar\sigma_m\tilde\lambda)\!+\!D_l(\eps\sigma_m\bar{\tilde{\lambda}}\!-\!\bar\eps\bar\sigma_{m}\tilde\lambda)\!                        -\!D_m(\eps\sigma_l\bar{\tilde{\lambda}}\!-\!\bar\eps\bar\sigma_{l}\tilde\lambda)\}  \Big],}\\
{\tilde\delta_{\eps} \tilde\lambda_{\alpha}=-\epsilon_{\alpha}\tilde d+ 2i\eps_\ga{(\s^{mn})^\g}_\a \tilde f_{mn}+
\frac{h}{4}\omega^{nl}\;\Big[-\!\frac{1}{4}\{\eps\sigma_n\bar{\tilde{\lambda}}\!-\!\bar\eps\bar\sigma_n\tilde\lambda,2D_l\tilde\lambda_{\alpha}\!-\!i[\tilde a_l,\tilde\lambda_{\alpha}]\}}\\
{\phantom{\tilde\delta_{\eps} \tilde\lambda_{\alpha}=}-i\eps_\ga{(\s^{mk})^\g}_\a
\big(4\{\tilde f_{mn},\tilde f_{kl}\}-2\{\tilde a_n,D_l \tilde f_{mk}+\partial_l \tilde f_{mk}\}\big)\!
}\\
{\phantom{\tilde\delta_{\eps} \tilde\lambda_{\alpha}=}\!-\!\{\tilde a_n,4iD_l(\eps_\ga{(\s^{mk})^\g}_\a \tilde f_{mk})
+\!2[\tilde a_l,\eps_\ga{(\s^{mk})^\g}_\a \tilde f_{mk}]\!+\!\frac{i}{4}[\eps\sigma_l\bar{\tilde{\lambda}}\!-\!\bar\eps\bar\sigma_l\tilde\lambda,\tilde\lambda_\a]\}\Big],}\\
{\tilde\delta_{\eps} \tilde d= i\epsb\sb^m D_m\tilde\lambda+i\eps\s^m D_m
\bar{\tilde{\lambda}}\;+\frac{h}{4}\omega^{nl}\,\Big[2i\{\tilde f_{mn},\bar\eps\bar\s^m D_l\tilde\la+\eps\s^m D_l\bar{\tilde\la}\} }\\
{\phantom{\tilde\delta_{\eps}\tilde d=}+i\{\tilde a_n,(\partial_l +D_l)(\bar\eps\bar\s^m D_m\tilde\la+\eps\s^m D_m\bar{\tilde\la})\}
\!-\!\frac{1}{4}\{\eps\sigma_n\bar{\tilde{\lambda}}\!-\!\bar\eps\bar\sigma_n\tilde\lambda,2D_l \tilde d\!-\!i[\tilde a_l,\tilde d]\}}\\
{\phantom{\tilde\delta_{\eps}\tilde d=}\!-\!\{\tilde a_n,2D_l(i\epsb\sb^m D_m\tilde\lambda+i\eps\s^m D_m
\bar{\tilde\lambda})\!-\!i[\tilde a_l,i\epsb\sb^m D_m\tilde\lambda+i\eps\s^m D_m
\bar{\tilde\lambda}]
\!+\!\frac{i}{4}[\eps\sigma_l\bar{\tilde{\lambda}}\!-\!\bar\eps\bar\sigma_l\tilde\lambda,\tilde d]\}\Big].}\\
\end{array}
\label{nonlinsusy}
\end{equation}
 We have thus worked out, up to first order in $h\omega^{mn}$, the infinitesimal variations of the ordinary fields that
give rise through the Seiberg-Witten map in eq.~\eqref{standardWSmap} to
the linearly realised supersymmetric transformations
--see eq.~\eqref{ncsusytransf}-- of the noncommutative fields. Of course, if
we set $h=0$, these infinitesimal variations of the ordinary fields boil down to the ordinary supersymmetry transformations of an ordinary gauge theory in the Wess-Zumino gauge. However, the contributions of order $h\omega^{mn}$ are
nonlinear modulo gauge transformations, and tell us that unlike for gauge
symmetries the standard Seiberg-Witten map in eq.~\eqref{standardWSmap} does not
transmute supersymmetry transformations of the ordinary fields realising supersymmetry linearly into supersymmetry transformations of the noncommutative fields also realising supersymmetry linearly. The question then arises as to whether
the nonlinear transformations in eq.~\eqref{nonlinsusy} realise a
--nonlinear-- representation of supersymmetry in the sense that the commutator
of two such transformations on ordinary fields closes on space-time
translations modulo ordinary gauge transformations.
If we can answer the question in the affirmative --which we shall, at any
order in  $h\omega^{mn}$--, we will be entitled to call the transformations
in eq.~\eqref{nonlinsusy} supersymmetry transformations. This issue has
never been discussed in the literature, although the $U(1)$ version of the transformations in eq.~\eqref{nonlinsusy} have been called supersymmetry transformations. Let us show that if $\tilde\delta_{\eps}{\it u}$ is an infinitesimal
transformation satisfying eq.~\eqref{ordsusyeq}, then
\begin{equation}
[\tilde\delta_\xi,\tilde\delta_\eta]{\it u}(x)=
-2i(\eta\sigma^m\bar\xi-\xi\sigma^m\bar\eta)\,\partial_m\;{\it u}(x)+
\delta^{(gauge)}_{g(x)}\;{\it u}(x)\equiv (P+\delta^{(gauge)}_{g(x)})\;{\it u}(x),
\label{ordsusyalg}
\end{equation}
where $g(x)$ is the inverse image of  $\Omega(x)$ in eq.~\eqref{closureNC} under the Seiberg-Witten map, i.e., --see eq.~\eqref{SWmapforZ}--
\begin{displaymath}
\Omega(x)=g(x)+\frac{h}{4}\omega^{mn}\,\{\tilde a_m,\partial_n g\}(x) + O(h^2).
\end{displaymath}
Now, since $\tilde\delta_\xi$ and $\tilde\delta_\eta$ are infinitesimal variations, their commutator $[\tilde\delta_\xi,\tilde\delta_\eta]$ acts as a derivation
on polynomials of the ordinary fields and their space-time derivatives. Then
\begin{displaymath}
[\tilde\delta_\xi,\tilde\delta_\eta]{\cal U}[\tilde a_m,{\it u}]=
{\cal U}[(1+[\tilde\delta_\xi,\tilde\delta_\eta])\tilde a_m,(1+[\tilde\delta_\xi,\tilde\delta_\eta]){\it u}]-{\cal U}[\tilde a_m,{\it u}]\,+\,\text{higher orders},
\end{displaymath}
where ${\cal U}[\tilde a_m,{\it u}]$ is the formal power series expansion that implements
the Seiberg-Witten map. Taking into account eq.~\eqref{ordsusyeq}, one concludes that
\begin{equation}
[\widehat\delta_\xi,\widehat\delta_\eta]\,{\cal U}[\tilde a_m,{\it u}]=
[\tilde\delta_\xi,\tilde\delta_\eta]\,{\cal U}[\tilde a_m,{\it u}]=
{\cal U}[(1+[\tilde\delta_\xi,\tilde\delta_\eta])\tilde a_m,(1+[\tilde\delta_\xi,\tilde\delta_\eta]){\it u}]-{\cal U}[\tilde a_m,{\it u}]\,+\,\text{higher orders}.
\label{commutatorone}
\end{equation}
On the other hand, eq.~\eqref{closureNC} leads to
\begin{equation}
\begin{array}{l}
{[\widehat\delta_\xi,\widehat\delta_\eta]\,{\cal U}[\tilde a_m,{\it u}]=
(P+\delta^{(ncgauge)}_{\Omega(x)})\,{\cal U}[\tilde a_m,{\it u}]=
(P+\delta^{(gauge)}_{g(x)})\,{\cal U}[\tilde a_m,{\it u}]}\\
{\phantom{[\widehat\delta_\xi,\widehat\delta_\eta]\,{\cal U}[\tilde a_m,{\it u}]}
={\cal U}[(1+P+\delta^{(gauge)}_{g(x)})\,\tilde a_m,(1+P+\delta^{(gauge)}_{g(x)})\,{\it u}]
-{\cal U}[\tilde a_m,{\it u}]\,+\,\text{higher orders},}
\label{commutatortwo}
\end{array}
\end{equation}
upon using the fact that by  definition of the Seiberg-Witten map we
have $\delta^{(ncgauge)}_{\Omega(x)}\,{\cal U}[\tilde a_m,{\it u}]=
 \delta^{(gauge)}_{g(x)}\,{\cal U}[\tilde a_m,{\it u}]$. Finally, eqs.~\eqref{commutatorone} and~\eqref{commutatortwo} imply that
 \begin{displaymath}
{\cal U}[(1+[\tilde\delta_\xi,\tilde\delta_\eta])\,\tilde a_m,(1+[\tilde\delta_\xi,\tilde\delta_\eta])\,{\it u}]=
{\cal U}[(1+P+\delta^{(gauge)}_{g(x)})\,\tilde a_m,(1+P+\delta^{(gauge)}_{g(x)})\,{\it u}],
\end{displaymath}
which in turn yields eq.~\eqref{ordsusyalg}. Let us stress that the two key
facts we have taken advantage of to obtain eq.~\eqref{ordsusyalg} are that
our noncommutative fields carry a  representation of the supersymmetry algebra
and that the Seiberg-Witten map turns (ordinary) gauge transformations of the ordinary fields into (noncommutative) gauge transformations of the noncommutative fields. Our proof of eq.~\eqref{ordsusyalg} is valid to all orders in powers
of $h\omega^{mn}$ and for any type of $U(N)$ Seiberg-Witten map provided
$\tilde\delta_{\eps}{\it u}(x)$ exists.

To close the current section let us remark that having a nonlinear realisation of the ${\cal N}=1$ supersymmetry algebra in four dimensions as furnished by 
the transformations in eq.~\eqref{nonlinsusy} is in keeping with the duality that seems to establish the standard Seiberg-Witten map --supplemented with a field 
redefinition-- between two supersymmetric  DBI actions in four dimensions, namely, the noncommutative $U(1)$ supersymmetric DBI action and the ordinary $U(1)$ supersymmetric DBI action in the presence of a background field $B_{mn}$. Indeed,we show in Appendix B that a given field redefinition of the Seiberg-Witten map in 
eq.~\eqref{standardWSmap} turns, for small $B_{mn}$ and up to order 4 in the
susy field strength, the ordinary 
$U(1)$ supersymmetric DBI action for a background field 
$B_{mn}$ in four dimensions into the leading contribution to the noncommutative $U(1)$ supersymmetric DBI action; the latter being the action of noncommutative $U(1)$ ${\cal N}=1$ superYang-Mills theory. Now, in four dimensions, the gauge supermultiplet of the ordinary $U(1)$ supersymmetric DBI theory in a background field $B_{mn}$, as formulated in ref.~\cite{Seiberg:1999vs},  carries a nonlinear realisation of the 
${\cal N}=1$ supersymmetry algebra which is an unbroken symmetry of the corresponding DBI action. This nonlinear realisation of the supersymmetry algebra is~\cite{Seiberg:1999vs} a $B_{mn}$-dependent linear combination of the extensions to the case of
nonvanishing $B_{mn}$ of the linear (unbroken)  and the nonlinear 
(broken) supersymmetry transformations that leave invariant the DBI action for $B_{mn}=0$ in four dimensions.

\section{Only one dual ordinary theory}

In Section 2, we constructed an ordinary $U(1)$ gauge theory whose fields
carry a linear  realisation of ${\cal N}=1$ supersymmetry in four dimensions
and is dual under the Seiberg-Witten map 
for superfields to noncommutative $U(1)$ ${\cal N}=1$ superYang-Mills.
The Seiberg-Witten map that connects these ordinary and noncommutative supersymmetric gauge theories is nonlocal --see eqs.~\eqref{SWV}-- but its nonlocal contributions are mere gauge artifacts. In Section 3, we used the standard
--local-- Seiberg-Witten map in the Wess-Zumino gauge to construct an ordinary dual of noncommutative $U(1)$ ${\cal N}=1$ superYang-Mills, the ordinary fields of this ordinary dual carrying a nonlinear realisation of
the ${\cal N}=1$ supersymmetry algebra in four dimensions. The standard
Seiberg-Witten map giving the latter ordinary dual of   noncommutative  $U(1)$ ${\cal N}=1$ superYang-Mills is given in 
eq.~\eqref{standardWSmap}. The purpose of the current section is to show, at first order in $h\omega^{mn}$, that the 
ordinary duals of  noncommutative $U(1)$ ${\cal N}=1$ superYang-Mills that we have constructed in Sections 2 and 3
are not different ordinary $U(1)$ supersymmetric gauge theories but, indeed,
the same ordinary theory each time formulated in terms of a
different set of field variables:
one set of fields represents the ${\cal N}=1$ supersymmetry algebra linearly
and the other set nonlinearly. Before we show this, we must change, as usual, the normalisation of the noncommutative, $(A_m,\Lambda_{\a},D)$, and ordinary, $(\tilde a_m,\tilde\lambda_{\a},\tilde d)$, gauge supermultiplets of
Section 3 so that their gauge transformations have the same normalisation
as the gauge transformations for components derived from  the superfield gauge
transformations used in Section 2. The
normalisation change in question is the following:
  $(A_m,\Lambda_{\a},D)\rightarrow (\frac{1}{2} A_m,\Lambda_{\a},D)$ and $(\tilde a_m,\tilde\lambda_{\a},\tilde d)\rightarrow (\frac{1}{2} \tilde a_m,\tilde \lambda_{\a},\tilde d)$. This
change of normalisation turns the the Seiberg-Witten map in eq.~\eqref{standardWSmap} into the following Seiberg-Witten map:
\begin{equation}
\begin{array}{l}
{A_m[\tilde a_n]=\tilde a_m+ h A^{(1)\,st}_m+O(h^2),\quad\quad\quad
A^{(1)\,st}_m=\frac{1}{2}\omega^{nl}\big(\tilde a_n\partial_l \tilde a_m-\frac{1}{2} \tilde a_n\partial_m \tilde a_l\big),}\\
{\Lambda_{\alpha}[\tilde a_m,\tilde\lambda_\alpha]=\tilde\lambda_{\alpha}+h \Lambda^{(1)\,st}_{\alpha}+O(h^2),\quad\quad\quad
\Lambda^{(1)\,st}_{\alpha}=\frac{1}{2}\omega^{mn}\,\tilde a_m\partial_n\tilde\lambda_{\alpha},}\\
{D[\tilde a_m,\tilde d]=\tilde d+
h D^{(1)\,st}+O(h^2),\quad\quad\quad\quad
D^{(1)\,st}=\frac{1}{2}\omega^{mn}\tilde a_m \partial_n \tilde d.}
\label{standardWSmapnorm}
\end{array}
\end{equation}
Let us next establish a map between
the ordinary gauge supermultiplet $(a_m,\lambda_\a,d)$ that occurs in the map
in eq.~\eqref{nonlocalWZmap}  and the ordinary gauge supermultiplet
$(\tilde a_m,\tilde\lambda_\a,\tilde d)$ that is in the Seiberg-Witten map in
eq.~\eqref{standardWSmapnorm}.
We shall first remind the reader that the map between the noncommutative supermultiplet
$(A_m,\Lambda_\a,D)$ and the ordinary supermultiplet $(a_m,\lambda_\a,d)$ defined by $V^{WZ}[a_m,\lambda_\a,d]$ in eq.~\eqref{nonlocalWZmap} is obtained by gauge transforming to the Wess-Zumino gauge --see eqs.~\eqref{gaugingtoWZgauge} to ~\eqref{nonlocalWZmap}--  the Seiberg-Witten map defined by
eqs.~\eqref{firstordexp} and~\eqref{SWV}, when $x=0$ and ${\cal X}=0$ --recall that ${\cal X}=0$ corresponds to an ordinary local field redefinition and therefore bears no physical consequences. Now, one may show that $A^{(1)}_{\beta\a}$,$\Lambda^{(1)}_\a$ and $D^{(1)}$ in eqs.~\eqref{V1components}~and~\eqref{nonlocalWZmap} can expressed as follows
\begin{align}
\nonumber A_{\dot\b\g}^{(1)}&=A_{\dot\b\g}^{(1)\,
st}-2\partial_{\dot\b\g}{\cal Z}+{\cal A}_{\dot\b\g},\,s{\cal
A}_{\dot\b\g}=0,\\
\label{fieldred}\Lambda^{(1)}_\r&=\Lambda_\r^{(1)\, st}+{\cal
L}_\rho,\,s{\cal L}_{\r}=0,\\
\nonumber D^{(1)}&=D^{(1)\, st}+{\cal D},\,s{\cal D}=0,
\end{align}
where $A_{\dot\b\g}^{(1)\,
st}$, $\Lambda_\r^{(1)\, st}$  and $D^{(1)\, st}$ are obtained from the
functions denoted with the same symbol in eq.~\eqref{standardWSmapnorm} by replacing $(\tilde a_m,\tilde\lambda_\a,\tilde d)$ with  $(a_m,\lambda_a,d)$. ${\cal Z}$ and the BRS trivial pieces ${\cal A}_{\dot\b\g},{\cal
L}$ and ${\cal D}$ are displayed next:
\begin{align}
\nonumber{\cal Z}=&-\frac{1}{128}\om^{\a\b}\Big(a^{\ad}_\a-\frac{\partial^{\ad}_\a}{\Box}\partial
a\Big)a_{\ad\b}-\frac{1}{128}\om^{\ad\bd}\Big(a^{\a}_{\ad}-\frac{\partial^{\a}_{\ad}}{\Box}\partial
a\Big)a_{\bd\a},\\
\nonumber{\cal A}_{\bd\g}=&\frac{-1}{256}\om^{\a\b}\Big[4\Big(a_\a^{\ad}-\frac{\partial_\a^{\ad}}{\Box}\partial
a\Big)\partial_{\bd\ga}\Big(a_{\ad\b}-\frac{\partial_{\ad\b}}{\Box}\partial
a\Big)-8\Big(a_\a^{\ad}-\frac{\partial_\a^{\ad}}{\Box}\partial
a\Big)\partial_{\ad\b}\Big(a_{\bd\ga}-\frac{\partial_{\bd\ga}}{\Box}\partial
a\Big)\\
\nonumber&\phantom{\frac{i}{256}\om^{\a\b}\Big[}-\frac{\partial_\a^{\ad}}{\Box}d\frac{\partial_{\ad\b}\partial_{\bd\g}}{\Box}d+2i\frac{\partial_\a^{\ad}\partial_{\bd\s}}{\Box}\la^\s
\frac{\partial_{\ad\b}\partial_{\g\sd}}{\Box}\lab^{\sd}\Big]+(c.c.)|_{\b\leftrightarrow\ga},\\
\label{cals}{\cal
L}_\r=&\frac{1}{128}\om^{\a\b}\Big[4\Big(a_\a^{\ad}-\frac{\partial_\a^{\ad}}{\Box}
\partial a\Big)\partial_{\ad\b}\la_\r+
2\frac{\partial_{\a}^{\ad}\partial_\s^{\rd}}{\Box}\la^\s\partial_{\ad\b}\Big(a_{\rd\r}-\frac{\partial_{\rd\r}}{\Box}
\partial a\Big)+
i \frac{\partial_{\a}^{\ad}\partial_\s^{\rd}}{\Box}\la^\s
\frac{\partial_{\ad\b}\partial_{\rd\r}}{\Box}d               \Big]\\
\nonumber+&\frac{1}{128}\om^{\ad\bd}\Big[4\Big(a_{\ad}^{\a}-\frac{\partial_{\ad}^{\a}}{\Box}
\partial a\Big)\partial_{\bd\a}\la_\r+
2\frac{\partial_{\ad}^{\a}\partial_\s^{\rd}}{\Box}\la^\s\partial_{\bd\a}\Big(a_{\rd\r}-\frac{\partial_{\rd\r}}{\Box}
\partial a\Big)
+
i \frac{\partial_{\ad}^{\a}\partial_\s^{\rd}}{\Box}\la^\s
\frac{\partial_{\bd\a}\partial_{\rd\r}}{\Box}d
 \Big],\\
\nonumber{\cal D}=&\frac{1}{128}\om^{\a\b}\Big[4\Big(a_\a^{\ad}-\frac{\partial_\a^{\ad}}{\Box}\partial
a\Big)\partial_{\ad\b}d+2\partial_\a^{\ad}\Big(a^{\s\rd}-\frac{\partial^{\s\rd}}{\Box}\partial
a\Big)\frac{\partial_{\ad\b}\partial_{\rd\s}}{\Box}d+\frac{\partial_\a^{\ad}\partial_\s^{\rd}}{\Box}\la^\s\partial_{\ad\b}\lab_{\rd}+\\
\nonumber&+\partial_\a^{\ad}\la^\s\frac{\partial_{\ad\b}\partial_{\sd\s}}{\Box}\lab^{\sd}\Big]+(c.c.).
\end{align}
We finally define the following maps between the ordinary gauge supermultiplets $(a_m, \lambda_\a,d)$ --linear-- and $(\tilde a_m,\tilde\lambda_\a,\tilde d)$ --nonlinear:
\begin{equation}
\begin{array}{l}
{\tilde a_m=a_m-2h\partial_m{\cal Z}[a]+h{\cal
A}_m[a,\la,d]+O(h^2),}\\
{\tilde \la_\a=\la+h{\cal L}_\a[a,\la,d]+O(h^2),\quad\quad  \tilde d=d+h{\cal
D}[a,\la,d]+O(h^2),}\\
\end{array}
\label{mapnotildetotilde}
\end{equation}
where ${\cal Z}$ and the BRS-closed functions ${\cal
A}_m$, ${\cal L}_\a$ and ${\cal D}$ are given in eq.~\eqref{cals} --see also eq.~\eqref{fieldred}.

Let us discuss some properties of the map in eq.~\eqref{mapnotildetotilde}.
First, for infinitesimal $U(1)$ transformations, it maps orbits of $(a_m,\lambda_\a,d)$ into orbits of $(\tilde a_m,\tilde \lambda_\a,\tilde d)$, and viceversa. Indeed, using eq.~\eqref{mapnotildetotilde}, one may show that
\begin{align*}
 \tilde s_{\tilde{z}}(\tilde a_m,\tilde\la_\a,\tilde
d)= s_z(\tilde a_m,\tilde\la_\a,\tilde
d),\quad\quad \tilde z=z+h s_z{\cal Z}[a_n],
\end{align*}
where $s_z$ denotes the $U(1)$ BRS operator acting on $(a_m,\lambda_\a,d)$:
$s_za_m=-2\partial_m z$, $s_z\lambda_\a=0$ and $s_z d=0$, and $\tilde s_{\tilde z}$ stands for the $U(1)$ BRS operator acting on $(\tilde a_m,\tilde \lambda_\a,\tilde d)$:
$\tilde s_{\tilde{z}}\tilde a_m=-2\partial_m \tilde z$,
$s_{\tilde{z}}\tilde\la_\a=0$ and $s_{\tilde{z}}\tilde d=0$. Secondly, the fact
that under ${\cal N}=1$ supersymmetry transformations the supermultiplet $(a_m,\lambda_\a,d)$ transforms  as follows
\begin{equation}
\begin{array}{l}
{ \delta_\eps a_m=\frac{1}{2}\eps \s_m
\lab-\frac{1}{2}\epsb\sb_m\la,\quad \delta_\eps \la_\a=-\eps_\a
d+i\eps_\ga{(\s^{mn})^\g}_\a f_{mn},}\\
{\delta_\eps d=i\epsb\sb^m\partial_m\la+i\eps\s^m\partial_m \lab,
\quad f_{mn}=\partial_m
a_n-\partial_n a_m,}
\label{susyWZ}
\end{array}
\end{equation}
and eq.~\eqref{mapnotildetotilde}, lead to
\begin{displaymath}
\delta_\eps(\tilde a_m,\tilde \la_\a,\tilde d)=
(\tilde\delta_\eps+\tilde s_{\tilde
z})(\tilde a_m,\tilde\la_\a,\tilde d),\quad\quad\tilde z={\rm
Re}(ih\bar\eps\bar\Psi^{(1)})+h s_{{\rm
Re}(ih\bar\eps\bar\Psi^{(1)})}{\cal Z}[a],
\end{displaymath}
where $\tilde\delta_\eps(\tilde a,\tilde\la,\tilde d)$ are the nonlinear supersymmetry transformations in eq.~\eqref{nonlinsusy}
for $U(1)$ fields after the rescaling $\tilde a_m\rightarrow\frac{1}{2}\tilde a_m$, and  $\Psi^{(1)}$ and ${\cal Z}[a]$ are given in eqs.~\eqref{V1components}
and \eqref{cals}, respectively. Hence, modulo gauge transformations, the linear
supersymmetry transformations --eq.~\eqref{susyWZ}-- of the gauge supermultiplet $( a_m,\la_\a,\ d)$ imply the nonlinear
supersymmetry transformations of the gauge supermultiplet $\tilde\delta_\eps(\tilde a,\tilde\la,\tilde d)$
as defined in eq.~\eqref{nonlinsusy}; and viceversa. Finally, if
$(\tilde a_m,\tilde\la_\a,\tilde d)$ and $( a_m,\la_\a,\ d)$ satisfy eq.~\eqref{mapnotildetotilde}, then both gauge
supermultiplets will have the same noncommutative supermultiplet image, $(A_m,\Lambda_\a,D)$, under the corresponding maps in
eqs.~\eqref{nonlocalWZmap} and~\eqref{standardWSmapnorm}:
\begin{equation}
\begin{array}{l}
{A_m=a_m+hA^{(1)}_m[a_n\lambda_\a,d]+O(h^2)=
\tilde a_m+hA^{(1)\,st}_m[\tilde a_n]+O(h^2),}\\
{\Lambda_\a=\la_\a+h\Lambda^{(1)}_\a[a_n\lambda_\b,d]+O(h^2)=
\tilde \la_\a+h\Lambda^{(1)\,st}_\a[\tilde a_n,\tilde\lambda_\b]+O(h^2),}\\
{D=d +h D^{(1)}[a_n\lambda_\a,d]+O(h^2)=
\tilde d+h D^{(1)\,st}[\tilde a_n,\tilde d]+O(h^2).}\\
\end{array}
\label{samencsusymultiplet}
\end{equation}
Eq.~\eqref{fieldred} helps to show the previous set of equalities. We have thus
shown that the supermultiplets $(\tilde a_m,\tilde\la_\a,\tilde d)$ and
$(a_m,\la_\a, d)$ define, up to first order in $h\omega^{mn}$ the same
$U(1)$ ordinary supersymmetric gauge theory with no matter fields. Notice
that eqs.~\eqref{samencsusymultiplet} imply that the action in terms of $(\tilde a_m,\tilde\la_\a,\tilde d)$ is equal to the
action in terms of $(a_m,\la_\a,d)$, if these gauge supermultiplets are related by eq.~\eqref{mapnotildetotilde}.

We have thus shown that the ordinary theories dual to noncommutative SYM found in Sections 2 and 3 are
not different theories but the same ordinary supersymmetric gauge theory formulated in each case in terms of a different set of field variables. The ordinary field variables introduced in Section 2 carry a linearly realised ${\cal N}=1$ supersymmetry and the
set of ordinary fields of Section 3 transforms nonlinearly under  ${\cal N}=1$ supersymmetry.


\section{Summary and Conclusions}
In Section 2,  we have found, at first order in $h\omega^{mn}$, the most general solution to the Seiberg-Witten map equations for
a noncommutative $U(1)$ vector superfield that is a polynomial in its ordinary counterpart,$v$, the chiral and antichiral projections
of the latter, $v_+$ and $v_-$,  and the susy covariant derivatives of them all; such polynomial being at most quadratic in
$v$, $v_+$ and $v_-$. These Seiberg-Witten maps are nonlocal, but their nonlocal parts are gauge artifacts  since they can be set to zero by choosing the supersymmetric Landau gauge. Furnished with this family of solutions to the $U(1)$
Seiberg-Witten map equations, we have obtained an ordinary dual under the Seiberg-Witten map of noncommutative SYM. This ordinary dual when formulated in terms of the ordinary fields considered in Section 2 has linearly realised supersymmetry. In Section 2, we have also shown
by explicit computation that the standard Seiberg-Witten map of ref.~\cite{Seiberg:1999vs} is never the $\theta\bar\theta$ component
of a vector superfield which is a polynomial in the corresponding ordinary vector superfield and its susy covariant derivatives.
In Section 3, we have obtained the ordinary duals under the generalisation of the standard Seiberg-Witten  map of ref.~\cite{Seiberg:1999vs} of noncommutative $U(N)$ gauge theory with ${\cal N}=1$ supersymmetry. These duals have been obtained by formulating the
noncommutative theory in the Wess-Zumino gauge. The noncommutative fields of our noncommutative theory carry a linear realisation of the ${\cal N}=1$ supersymmetry algebra in four dimensions; however, as we have shown in Section 3, their ordinary counterparts under the standard Seiberg-Witten map carry a nonlinear representation of  the ${\cal N}=1$ supersymmetry algebra in four dimensions. Hence, the ordinary dual
of our noncommutative supersymmetric theory supports a nonlinear realisation of the supersymmetry algebra when formulated in terms of the
ordinary supermultiplets of Section 3. We have seen that this is in line with the 
duality under the Seiberg-Witten map --see Appendix B-- between the 
noncommutative $U(1)$ supersymmetric DBI theory and the
ordinary abelian supersymmetric DBI theory in a $B_{mn}$ field in four 
dimensions. In section 4, we have shown that the ordinary duals of noncommutative SYM constructed in Sections 2 
and 3 by using completely different types  of Seiberg-Witten map are not different ordinary supersymmetric gauge theories, but the same
ordinary theory formulated, in each case, in terms of a different set of field variables: a set of field variables carries a linear representation of ${\cal N}=1$ supersymmetry algebra in four dimensions and the other set carries a nonlinear representation of this algebra.
We define, in Section 4, the map that realises the change of field variables and study the properties of the map: it maps infinitesimal gauge orbits into infinitesimal gauge orbits and turns the linear
realisation of ${\cal N}=1$ supersymmetry in Section 2 into the $h\omega^{mn}$-dependent nonlinear realisation  of the latter in Section 3.

We believe that the results we have obtained in Sections 2 and 4 for $U(1)$ can be extended to $U(N)$ groups in the fundamental, antifundamental, adjoint and bifundamental representations. However, to obtain explicit expressions such as the Seiberg-Witten map
for superfields in eqs.~\eqref{SWV} will be much harder since the r.h.s. in eq.~\eqref{SWeq}
contains an infinite number of terms for nonabelian ordinary groups. We also believe that the results obtained in section 2 can be
extended to any ordinary nonabelian gauge group in any representation, if one adopts the general philosophy behind
the formalism put forward in refs.~\cite{Madore:2000en, Jurco:2000ja, Jurco:2001rq} for non-supersymmetric gauge theories: now the noncommutative vector superfields will be valued
in the enveloping algebra of the Lie algebra of the ordinary gauge group. Section 3, however, will not hold, in general, for a given ordinary gauge group in a given representation, e.g.,  $SU(N)$ in the fundamental representation. Indeed, generally speaking
$\tilde\delta_\eps \tilde a_m$ as defined in eq.~\eqref{nonlinsusy}   is not valued in the Lie algebra of the gauge group, so it is not, in general, a variation of an ordinary gauge field. It so happens that for arbitrary gauge groups in arbitrary representations, if the
enveloping-algebra-valued noncommutative fields of the gauge triplet $(A_m,\Lambda_\a,D)$ are defined in terms of  ordinary fields by means
of the standard Seiberg-Witten map, the linear supersymmetry transformations in eq.~\eqref{ncsusytransf}
are not given rise to by variations of the ordinary fields. In view of the important
results --see refs.~\cite{Calmet:2001na,Aschieri:2002mc}-- achieved within the enveloping-algebra formalism of refs.~\cite{Madore:2000en, Jurco:2000ja} and ~\cite{Jurco:2001rq}, it is worth exploring how to construct supersymmetric versions of the models in refs.~\cite{Calmet:2001na} and~\cite{ Aschieri:2002mc}. Perhaps, one should
look for $h\omega^{mn}$-dependent nonlinear realisations of supersymmetry carried by ordinary fields that yield upon using the standard
Seiberg-Witten map noncommutative fields that also carry an $h\omega^{mn}$-dependent nonlinear realisation of supersymmetry. Let us
notice that we cannot start with an ordinary gauge supermultiplet having standard linear supersymmetry transformations and then apply the standard Seiberg-Witten map to define the noncommutative fields, since, as we show in Appendix C, the ordinary action dual to the action
of noncommutative $U(1)$ gauge theory cannot be made supersymmetric under those linear supersymmetry transformations by adding local terms which are polynomials in $h\omega^{mn}$. Finally, perhaps, to generalise the formalism of  refs.~\cite{Madore:2000en, Jurco:2000ja}
and~\cite{Jurco:2001rq} so as to include supersymmetry, one should use the ideas and techniques in ref.~\cite{Dimitrijevic:2007cu}.

\section*{Acknowledgements}
This work has been financially supported in part by MEC through grants
FIS2005-02309 and PCI2005-A7-0153. The work of C. Tamarit has also received 
partial financial support from MEC trough FPU grant AP2003-4034. We would like
to thank Professors R. Banerjee and S. Ghosh for long and deep discussions 
on the issues addressed in this paper. We would also like to thank Professor
A. Tseytlin and Dr. M. Lledo for instructing us on some properties of the 
ordinary and noncommutative DBI actions.

\section{Appendix A. Superspace conventions}

       Our superspace conventions are those of
ref.~\cite{FigueroaO'Farrill:2001tr}. The superspace coordinates are
given by
$x^m,\th_\a,\bar\th_{\dot\a}$, with  $\bar\th_{\dot\a}=\th^\star_\a$.
We denote space-time indices with latin letters and spinor indices
with greek letters. Spinor indices are raised with and lowered with
$\epsilon^{\a\b},\epsilon_{\a\b},\epsilon^{\dot\a\dot\b},\epsilon_{\dot\a\dot\b}$
such that $\eps^{12}=1=\eps_{12}=-\eps^{\dot 1\dot 2}=-\eps_{\dot 1
\dot 2}$ and $\eps_{\a\b}^*=\eps_{\dot\b\dot\a}$. Contractions will be
denoted as $\eps\eta\equiv\eps^\a\eta_\a,
\bar\eps\bar\eta\equiv\bar\eps^{\dot\a}\bar\eta_{\dot\a}$. For the
sigma matrices we have
\begin{align*}
 (\sigma^m)^{\a\dot\a}=&(1,\vec\sigma),&(\bar\sigma^m)_{\dot\a\a}&=(-1,\vec\sigma),\\
 (\sigma^{mn}{)^\a}_\b=&\frac{1}{2}(\sigma^m\bar\sigma^n-\sigma^n\bar\sigma^m{)^\a}_\b,
& {(\bar\sigma^{mn})_{\dot\a}}^{\dot\b}=&\frac{1}{2}(\bar\sigma^m\sigma^n-\bar\sigma^n\bar\sigma^m{)_{\dot\a}}^{\dot\b}.
\end{align*}
       Superfields are functions over the superspace. We denote
noncommutative superfields with capital letters and ordinary
superfields with lower-case letters. An ordinary superfield $\chi$
transforms under supersymmetry as
\begin{align*}
\delta_\eps\chi(x,\th_\a,\bar\th_{\dot\a})=(-\epsilon
Q-\bar{\epsilon}\bar{Q})\chi(x,\th_\a,\bar\th_{\dot\a}),
\end{align*}
and identically for a noncommutative superfield $\Xi$. The generators $Q_\a,\bar Q_{\dot\a}$ satisfy the
supersymmetry algebra $\{Q_\a,\bar
Q_{\dot\a}\}=2i\bar{\sigma}_{\dot\a\a}^m\partial_m$; explicitly
\begin{align*}
 Q_\a=&\partial_\a+i\bar\th^{\dot\a}(\bar\sigma^m)_{\dot\a\a}\partial_m,
&  \bar Q_{\dot\a}=&\bar\partial_{\dot\a}+i(\bar\sigma^m)_{\dot\a\a}\th^\a\partial_m.
\end{align*}
The supersymmetric covariant derivatives $D_\a,\bar D_{\dot\a}$, which
satisfy $\{D_\a,Q_\b\}=0=\{D_\a,\bar Q_{\dot\b}\}=\{\bar D_
{\dot\a},Q_\b\}=\{\bar D_{\dot\a},\bar Q_{\dot\b}\}$ and
$\{D_\a,D_{\dot\a}\}=-2i\bar{\sigma}_{\dot\a\a}^m\partial_m$, are
\begin{align*}
 D_\a=&\partial_\a-i\bar\th^{\dot\a}(\bar\sigma^m)_{\dot\a\a}\partial_m,
&  \bar D_{\dot\a}=&\bar\partial_{\dot\a}-i(\bar\sigma^m)_{\dot\a\a}\th^\a\partial_m.
\end{align*}
 We consider the following component expansion of a real superfield $v$:
\begin{align*}
 \nonumber v(x,\th,\thb)=&c(x)+\theta^\a\psi_\a(x)+\thb^{\ad}
\psib_{\ad}(x)+\frac{1}{2}\th^2
f(x)+\frac{1}{2}\thb^2\bar{f}(x)+\th^\a\sb^m_{\bd\a}\thb^{\bd}
a_m+\frac{1}{2}\th^2\thb^{\ad} \lab'_{\ad}+\\
 \nonumber &\frac{1}{2}\thb^2\th^{\a} \la'_{\a}+\frac{1}{4}\thb^2\th^2 d',\\
\la'_\a\equiv&\la_\a-i\s^m_{\bd\a}\partial_m\psib^{\bd},d'\equiv d+\Box c,
\end{align*}
and similarly for a noncommutative real superfield $V$.


\section{Appendix B. Duality between noncommutative and ordinary supersymmetric $U(1)$ DBI theories}

The aim of this appendix  is to show the equivalence of the effective supersymmetric DBI actions for open strings ending on D-branes obtained, on the one hand, in noncommutative space-time, and on the other, in ordinary space-time but in the presence of a constant background $B_{mn}$. The first type of DBI actions have a linearly realised supersymmetry in terms of the noncommutative fields, while the ordinary DBI actions with a $B_{mn}$ background are invariant under non-linear supersymmetry transformations. The equivalence is provided by the Seiberg-Witten maps; this provides a natural understanding of the fact that ordinary fields in local SW maps always seem to transform non-linearly under supersymmetry.

 In the non-supersymmetric U(1) case, the equivalence was first 
noted by Seiberg and Witten \cite{Seiberg:1999vs}, and it was shown to be exact. In the supersymmetric case, both the ordinary and noncommutative actions are known ---see refs.~\cite{Bagger:1996wp,Grandi:2000gi}--- but their possible equivalence has not been studied. 
	Here we will show the equivalence in the limit of $h\omega\rightarrow0$ and for small values of the fields. We choose the 
$h\omega^{mn}\rightarrow0$ and not the Seiberg-Witten limit $\alpha'\rightarrow0$ because in the supersymmetric case the $\alpha'\rightarrow0$ limit requires a complicated reexpansion of the action, while the $h\omega^{mn}\rightarrow0$ limit is compatible with a perturbative definition of the DBI actions in terms of an expansion in the number of fields.
	Our aim is to show the equivalence of the DBI actions at first order in $h$ and up to products of three ordinary fields. The 
noncommutative DBI lagrangian, which we shall denote as $\hat{\cal L}_{DBI}$, is a functional of the noncommutative supersymmetric field strengths $\hat W_\a=-\frac{1}{4} \bar{D}^2(e_\star^{-V}\star D_\a e_\star^V)$. It is given by a sum of terms with even powers of $\hat W^2,\hat{\bar W}^2$~\cite{Grandi:2000gi}, so that it involves sums of products of an even number of component fields. We want to expand this action in terms of ordinary fields at first order in $h$ using the standard SW maps of eq.~\eqref{standardWSmapnorm}.

      	It can be easily seen that in order to compute the contributions with products of three ordinary fields and less, we only need $\hat{\cal L}_{DBI}$ up to $O(\hat W^2)$. Thus, following ~\cite{Grandi:2000gi} ---see \cite{Seiberg:1999vs} for the normalisation--- for a D3 brane we get,
\begin{align}
{\cal\hat L}_{DBI}=\frac{1}{2\pi G_s}\Big(\frac{1}{16}\int d^2\theta \hat W^2+\frac{1}{16}\int d^2\bar \theta \hat{\bar W}^2 \Big)+O(\hat W^4),
\label{NCDBI}
\end{align}
where $G_s$ is the noncommutative string coupling constant. In the component field expansion of $\hat W$ one must use the noncommutative space-time metric $G$.

	On the other hand, concerning the ordinary DBI action in the presence of the background field $B_{mn}$ ---which we shall denote as ${\cal L}_{DBI}$--- it is constructed from the 
action with $B_{mn}=0$ by making the substitution $f_{mn}\rightarrow f_{mn}-2B_{mn}$ ---the differences with the conventions in \cite{Seiberg:1999vs} are due to our choice of the component field expansion of the superfield $v$. The action at $B_{mn}=0$ is given by an expansion involving even powers of $W^2$, where $W_\a=-\frac{1}{4}D^2D_\a v$ is the ordinary supersymmetric field-strength, so that to get the terms with three fields after the substitution $f_{mn}\rightarrow f_{mn}-2B_{mn}$ we need the terms of ${\cal L}_{DBI}^{B=0}$ up to $O(W^4)$. These are given, adapting the result in \cite{Bagger:1996wp} to our conventions, by the following expression
\begin{align}
 {\cal L}_{DBI}=\frac{1}{2\pi g_s}\Big(\frac{1}{16}\int d^2\theta  W^2+\frac{1}{16}\int d^2\bar \theta {\bar W}^2 +\frac{(2\pi\alpha')^2}{128}\int d^2\theta d^2\bar\theta W^2\bar W^2+O(W^6)\Big)\Big|_{f\rightarrow f-2B}.
\label{DBI}
\end{align}
$g_s$ is the ordinary string coupling constant, and the ordinary metric $g$ must be used in the component field expansion of $W$.

	In order to relate both of the actions \eqref{NCDBI} and \eqref{DBI} in the limit of small $h\omega$, we need the results 
from \cite{Seiberg:1999vs} that follow
\begin{align}
 \frac{1}{G_s}=\frac{1}{g_s}+O(h^2),\quad G^{mn}=g^{mn}+O(h^2),\quad B=\frac{-1}{(2\pi\alpha')^2}g^{-1}h\omega g^{-1}+O(h^2).
\label{limit}
\end{align}
	For simplicity we can take both $G$ and $g$ as the Minkowski metric. We must expand both of the actions \eqref{NCDBI} and 
\eqref{DBI} in terms of the ordinary component fields and compare the results. Using the SW maps in \eqref{standardWSmapnorm}, the noncommutative action $\hat{\cal L}_{DBI}$ is given by
\begin{align}
 \nonumber\hat{\cal L}_{DBI}=&\frac{1}{2\pi g_s}\Big[-\frac{1}{16}f_{mn}f^{mn}+\frac{i}{16}\bar\lambda\bar\sigma^m\partial_m\lambda+\frac{1}{32}d^2-\frac{h}{64}\omega^{kl}f_{kl}f_{ij}f^{ij}+\frac{h}{16}\omega^{kl}f_{ik}f_{jl}f^{ij}\\
\label{NCDBIexp}&+\frac{ih}{128}\omega^{kl}f_{kl}(\bar\lambda\bar\sigma^m\partial_m\la-\bar\partial_m\lambda\bar\sigma^m\la)+\frac{ih}{64}\omega^{kl}f_{mk}(\bar\lambda\bar\sigma^m\partial_l\la-\bar\partial_l\lambda\bar\sigma^m\la)+\frac{h}{128}\omega^{kl}f_{kl}d^2
\Big]\\
\nonumber&+O(4\text{ fields})+O(h^2)+\text{total derivative}.
\end{align}
	The ordinary ${\cal L}_{DBI}$ action in eq.~\eqref{DBI} has the following component expansion, after using the relation 
between $\omega$ and $B$ in eq.~\eqref{limit}:
\begin{align*}
 {\cal L}_{DBI}=&\frac{1}{2\pi g_s}\Big[-\frac{1}{16}f_{mn}f^{mn}+\frac{i}{16}\bar\lambda\bar\sigma^m\partial_m\lambda+\frac{1}{32}d^2-\frac{h}{64}\omega^{kl}f_{kl}f_{ij}f^{ij}+\frac{h}{16}\omega^{kl}f_{ik}f_{jl}f^{ij}\\
&-\frac{ih}{256}\omega^{kl}f_{kl}(\bar\lambda\bar\sigma^m\partial_m\la
-\bar\partial_m\lambda\bar\sigma^m\la)+\frac{h}{256}\tilde\omega^{kl}f_{kl}\partial_m(\bar\la\bar\sigma^m\la)\\
&+\frac{ih}{128}\omega^{kl}f_{mk}(\bar\lambda\bar\sigma_l\partial^m\la-\bar\partial^m\lambda\bar\sigma_l\la)+\frac{ih}{128}\omega^{kl}f_{mk}(\bar\lambda\bar\sigma^m\partial_l\la-\bar\partial_l\lambda\bar\sigma^m\la)\\
&-\frac{h}{128}\epsilon^{lmqt}\omega_{kl}{f_m}^k\partial_q(\bar\la\bar\sigma_t\la)-\frac{h}{256}\omega^{kl}d\partial_l(\bar\la\bar\sigma_k\la)+\frac{ih}{256}\tilde\omega^{kl}d(\partial_k\bar\la\bar\sigma_l\la-\bar\la\bar\sigma_l\partial_k\la)\\
&-\frac{h}{128}\omega^{kl}f_{kl}d^2
\Big]
+O(4\text{ fields})+O(h^2)+\text{total derivative},
\end{align*}
where we have defined $\tilde\omega^{kl}\equiv\frac{1}{2}\epsilon^{klmn}\omega_{mn}$. At first sight, it is clear that the terms involving $f_{mn}$ alone coincide, as is known from previous results concerning non-supersymmetric theories. Still, the rest of the terms do not seem to match. However, we must still note that the SW maps are not uniquely defined, since they have an ambiguity given, in the U(1) case, by field redefinitions. Hence, we should check whether redefining the fields in the lagrangian ${\cal L}_{DBI}$ we can exactly match $\hat{\cal L}_{DBI}$ of eq.~\eqref{NCDBIexp}. The answer turns out to be positive in a non-trivial way. Indeed, it can be seen after some work that the following field redefinitions
\begin{align*}
 \delta a_m&=\frac{h}{16}{{\tilde\omega}_m}^{\,\,\,\,\,n}\bar\la\bar\sigma_n\la,\\
\delta \la_\a&=-\frac{3ih}{16}\tilde{\omega}^{kl}f_{kl}\la_\a+\frac{3h}{16}\omega^{kl}f_{kl}\la_\a+\frac{h}{8}\omega^{kl}{f^m}_k(\sigma_{lm})^{\a\b}\la_\b+\frac{ih}{4}\tilde\omega^{kl}{f^m}_k(\sigma_{lm})^{\a\b}\la_\b,\\
\delta d&=\frac{h}{4}\omega^{kl}f_{kl}d-\frac{h}{16}\omega^{kl}\partial_k(\bar\la\bar\sigma_l\la)-\frac{ih}{16}\tilde{\omega}^{kl}(\partial_k\bar\la\bar\sigma_l\la-\bar\la\bar\sigma_l\partial_k\la).
\end{align*}
turn ${\cal L}_{DBI}$ into $\hat{\cal L}_{DBI}$, modulo total derivatives and working at order $h$ and with terms involving products of up to three component fields. This is not trivial since even when considering the previous field redefinitions with arbitrary coefficients for the different terms, one cannot generate in the action ${\cal L}_{DBI}$ the terms appearing in $\hat{\cal L}_{DBI}$ with arbitrary coefficients. This shows that both DBI actions are in fact equivalent at least in the limit of small $h\omega^{mn}$ and small values of the fields, and this equivalence is provided by the Seiberg-Witten map in eq.~\eqref{standardWSmapnorm} supplemented with the previous field redefinitions. I.e., the modified Seiberg-Witten maps that follow,
\begin{align*}
A_m&=a_m+\frac{h}{2}\omega^{kl}\big(a_k\partial_l a_m-\frac{1}{2}a_k\partial_m a_l\big)-\frac{h}{16}{{\tilde\omega}_m}^{\,\,\,\,\,n}\bar\la\bar\sigma_n\la+O(h^2),\\
\Lambda\!&=\!\lambda+\frac{h}{2}\omega^{kl}a_k\partial_l\lambda+\frac{3ih}{16}\tilde{\omega}^{kl}f_{kl}\la_\a\!-\!\frac{3h}{16}\omega^{kl}f_{kl}\la_\a-\frac{h}{8}\omega^{kl}{f^m}_k(\sigma_{lm})^{\a\b}\la_\b-\frac{ih}{4}\tilde\omega^{kl}{f^m}_k(\sigma_{lm})^{\a\b}\la_\b\\
&+O(h^2),\\
D&=d+\frac{h}{2}\omega^{kl}a_k\partial_ld- \frac{h}{4}\omega^{kl}f_{kl}d+\frac{h}{16}\omega^{kl}\partial_k(\bar\la\bar\sigma_l\la)+\frac{ih}{16}\tilde{\omega}^{kl}(\partial_k\bar\la\bar\sigma_l\la-\bar\la\bar\sigma_l\partial_k\la)+O(h^2). 
\end{align*}
map $\hat{\cal L}_{DBI}$ of eq.~\eqref{NCDBI} into the action ${\cal L}_{DBI}$ of eq.~\eqref{DBI}.

	It is worth noting that, in the pure bosonic case, there is no need to consider field redefinitions; in fact the 
equivalence of the pure bosonic parts of $\hat{\cal L}_{DBI}$ and ${\cal L}_{DBI}$ was shown to be exact without having to use field redefinitions. This is due to the fact that, at least at order $h$ and possibly beyond, the pure bosonic field redefinitions only modify the bosonic lagrangian with pure derivative terms, so that their effect can be neglected.


\section{Appendix C: Is there a local linear supersymmetric completion of the
bosonic Yang-Mills action expanded with the standard SW map?}

       In Section 1 it was shown that the standard SW map can never
be embedded into a superfield. Furthermore, we have
seen that when considering local SW maps in components, the ordinary
fields transform in a non-linear representation of the supersymmetry
algebra. In all these cases, it was assumed that supersymmetry was
linearly realised on the side of the noncommutative fields. However,
there is still the possibility of the ordinary fields being in a
linear representation of supersymmetry and the noncommutative ones in
a non-linear one. We can thus start assuming a linear representation
of supersymmetry on the WZ gauge component fields $a_m,\la,d$, i.e.,
they should transform as in eq.~\eqref{susyWZ}. With this point of
view, the transformation properties of the noncommutative fields are
unknown and so is the action in terms of noncommutative fields.
Nevertheless, we know its pure bosonic part, which is the
noncommutative Yang-Mills expanded with the SW map.
Assuming further that the standard SW map \eqref{Ast} is valid for the
$A_m$ component, we have that the bosonic part of the action is given
by
\begin{align}
 \label{Sbosonicexp}S_{\rm bosonic}=&-\frac{1}{16}\idx F_{mn}\star F^{mn}=\\
&-\frac{1}{16}\idx f_{mn} f^{mn}-\frac{h}{64}\idx\,
\omega^{ab}f_{ab}f_{mn}f^{mn}+\frac{h}{16}\idx\,\omega^{ab}f_{ma}f_{nb}f^{mn}+O(h^2),
\nonumber
\end{align}
where the awkward normalisation factors are due to our unconventional
definitions of the component fields $A_m,a_m$. What needs
to be checked is whether there is any local, Poincar\'e and gauge
invariant  completion
of the action \eqref{Sbosonicexp} involving the WZ component fields
$a_m,\la,d$ which is invariant under the supersymmetric
transformations of eq.~\eqref{susyWZ}. Since the order $O(h^0)$ part
is known to have a supersymmetric completion, it suffices to check the
$O(h)$ part. To do so we consider all the possible independent
---modulo integration by parts--- gauge invariant monomials which are
of order one in $\omega^{mn}$, constructed from the fields $a_m,\la,d$ and
spacetime derivatives, which include at least one superpartner field
$\la,d$. They are shown next:
\begin{align*}
 \begin{array}{lll}
 t_1=\omega^{mn}f_{mr}{f_n}^rD,     & t_2=\omega^{mn}f_{mn}d^2, &
t_3=\omega^{mn}\partial_r f_{mn}\la\sigma^r\bar\la,\\
 t_4= \omega^{mn}\partial^rf_{mr}\la\sigma_n\bar\la,   &
t_5=\omega^{mn}\partial_nd\la\sigma_m\bar\la,  &
t_6=\omega^{mn}f_{mn}\square d, \\
 t_7={\rm Im}\, \omega^{mn}f_{mn}\la\prslash\bar\la,    & t_8={\rm
Im}\, \omega^{mn}f_{mr}\la\sigma_n\partial^r\bar\la,   & t_9={\rm Im}\,
\omega^{mn}f_{mr}\la\sigma^r\partial_n\bar\la, \\
t_{10}={\rm Im}\, \omega^{mn}d\la\sigma_m\partial_n\bar\la,     &
t_{11}={\rm Im}\, \omega^{mn}\square\partial_m\la\sigma_n\bar\la.\\
\end{array}
\end{align*}
``Im'' denotes imaginary part. By solving
\begin{align*}
 \widehat\delta_\eps\Big[S^{\rm bosonic}+h\idx\sum_i\alpha_i t_i\Big]=0
\end{align*}
expanding the l.h.s. in integrals of independent monomials, one readily
finds that there is no solution to the previous equation. This can be
seen for example by considering just the terms of the type
$ff\la,ff\bar\la$, which are the only ones generated from the
supersymmetric variation of the $fff$ terms of the bosonic action, as
is clear from eq.~\eqref{susyWZ}.

       Thus, the noncommutative Yang-Mills action expanded with the standard
SW map has no completion invariant under the linear
supersymmetry from eq.~\eqref{susyWZ}.

\end{document}